\RequirePackage{fix-cm}
\documentclass[twocolumn,epjc3]{svjour3}  
\smartqed  
\RequirePackage{graphicx}
\usepackage{color}
\usepackage{ulem}
\journalname{Eur. Phys. J. C}
\begin{document}

\title{Simulation Software of the JUNO Experiment 
}


\author{
Tao Lin\thanksref{IHEP} \and
Yuxiang Hu\thanksref{IHEP,ucas} \and
Miao Yu\thanksref{WHU} \and
Haosen Zhang\thanksref{IHEP,ucas} \and
Simon Charles Blyth\thanksref{IHEP} \and
Yaoguang Wang\thanksref{IHEP} \and
Haoqi Lu\thanksref{IHEP} \and
Cecile Jollet\thanksref{LP} \and
João Pedro  Athayde Marcondes de André\thanksref{IPHC} \and
Ziyan Deng\thanksref{e1,IHEP} \and
Guofu Cao\thanksref{e2,IHEP,ucas,keylab} \and
Fengpeng An\thanksref{SYSU} \and
Pietro Chimenti\thanksref{UEL} \and
Xiao Fang\thanksref{IHEP,xfang} \and
Yuhang Guo\thanksref{XJTU} \and
Wenhao Huang\thanksref{SDU} \and
Xingtao Huang\thanksref{SDU} \and
Rui Li\thanksref{SJTU} \and
Teng Li\thanksref{SDU} \and
Weidong Li\thanksref{IHEP} \and
Xinying Li\thanksref{IHEP,xyli} \and
Yankai Liu\thanksref{XJTU} \and
Anselmo  Meregaglia\thanksref{LP} \and
Zhen Qian\thanksref{SYSU} \and
Yuhan Ren\thanksref{IHEP,ucas} \and
Akira Takenaka\thanksref{SJTU} \and
Liangjian Wen\thanksref{IHEP} \and
Jilei Xu\thanksref{IHEP} \and
Zhengyun You\thanksref{SYSU} \and
Feiyang Zhang\thanksref{SJTU} \and
Yan Zhang\thanksref{IHEP,yzhang} \and
Yumei Zhang\thanksref{SYSU} \and
Jiang Zhu\thanksref{SYSU} \and
Jiaheng Zou\thanksref{IHEP}
}

\newcommand\mycolor{black}

\thankstext{e1}{e-mail: dengzy@ihep.ac.cn (corresponding author)}
\thankstext{e2}{e-mail: caogf@ihep.ac.cn (corresponding author)}


\institute{Institute of High Energy Physics, Beijing 100049, China \label{IHEP}
           \and
           University of Chinese Academy of Sciences, Beijing 100049, China \label{ucas}
           \and
           Wuhan University, Wuhan 430072, China \label{WHU}
           \and
           University of Bordeaux, CNRS, LP2i, Bordeaux, France \label{LP}
           \and
           Institut Pluridisciplinaire Hubert Curien, Université de Strasbourg, Strasbourg, France \label{IPHC}
           \and
           State Key Laboratory of Particle Detection and Electronics, Beijing 100049, China \label{keylab}
           \and
           Sun Yat-Sen University, Guangzhou 510275, China \label{SYSU}
           \and
           Universidade Estadual de Londrina, Londrina, Brazil \label{UEL}
           \and
           Xi'an Jiaotong University, Xi'an 710049, China \label{XJTU}
           \and
           Shandong University, Qingdao 266237, China \label{SDU}
           \and
           Shanghai Jiao Tong University, Shanghai 200240, China \label{SJTU}
           \and
           Present Address: Chengdu Documentation and Information Center, Chinese Academy of Sciences, Chengdu, China. \label{xfang}
           \and
           Present address: Neusoft Medical Systems Co., Ltd, Shenyang, China. \label{xyli}
           \and
           Present address: China Telecom Corporation Limited, Beijing, China. \label{yzhang}
}

\date{Received: date / Accepted: date}

\maketitle

\begin{abstract}
The Jiangmen Underground Neutrino Observatory (JUNO) is a multi-purpose experiment, under construction in southeast China, that is designed to determine the neutrino mass ordering and precisely measure neutrino oscillation parameters. Monte Carlo simulation plays an important role for JUNO detector design, detector commissioning, offline data processing, and physics processing. {\color{\mycolor}{The JUNO experiment has the world’s largest liquid scintillator detector instrumented with many thousands of PMTs. The broad energy range of interest, long lifetime, and the large scale present data processing challenges across all areas.}} This paper describes the JUNO simulation software, highlighting the challenges of JUNO simulation and solutions to meet these challenges, including such issues as support for time-correlated analysis, event mixing, event correlation and handling the simulation of many millions of optical photons.   
\end{abstract}


\section{Introduction}%
The JUNO experiment (Jiangmen Underground Neutrino Observatory)~\cite{JUNO:2015zny,JUNO:2022hxd} is a multi-purpose experiment designed to determine the neutrino mass ordering, precisely measure neutrino oscillation parameters, and probe the fundamental properties of neutrinos by detection of solar neutrinos, galactic core-collapse supernova neutrinos, atmospheric neutrinos, and geo-neutrinos. JUNO will also implement a dedicated multi-messenger (MM) trigger system maximizing its potential as a neutrino telescope by providing sensitivity to low-energy events. In addition, JUNO will provide an excellent environment for nucleon decay searches. JUNO is currently under construction near the city of Jiangmen in southern China, at a distance of 53 km from the Yangjiang and Taishan nuclear power plants, in an underground site with an overburden of $\sim$650~m (1800 m.w.e.). Detector commissioning is expected to start in 2023.

\begin{figure}[htb]\centering
	\includegraphics[width=85mm]{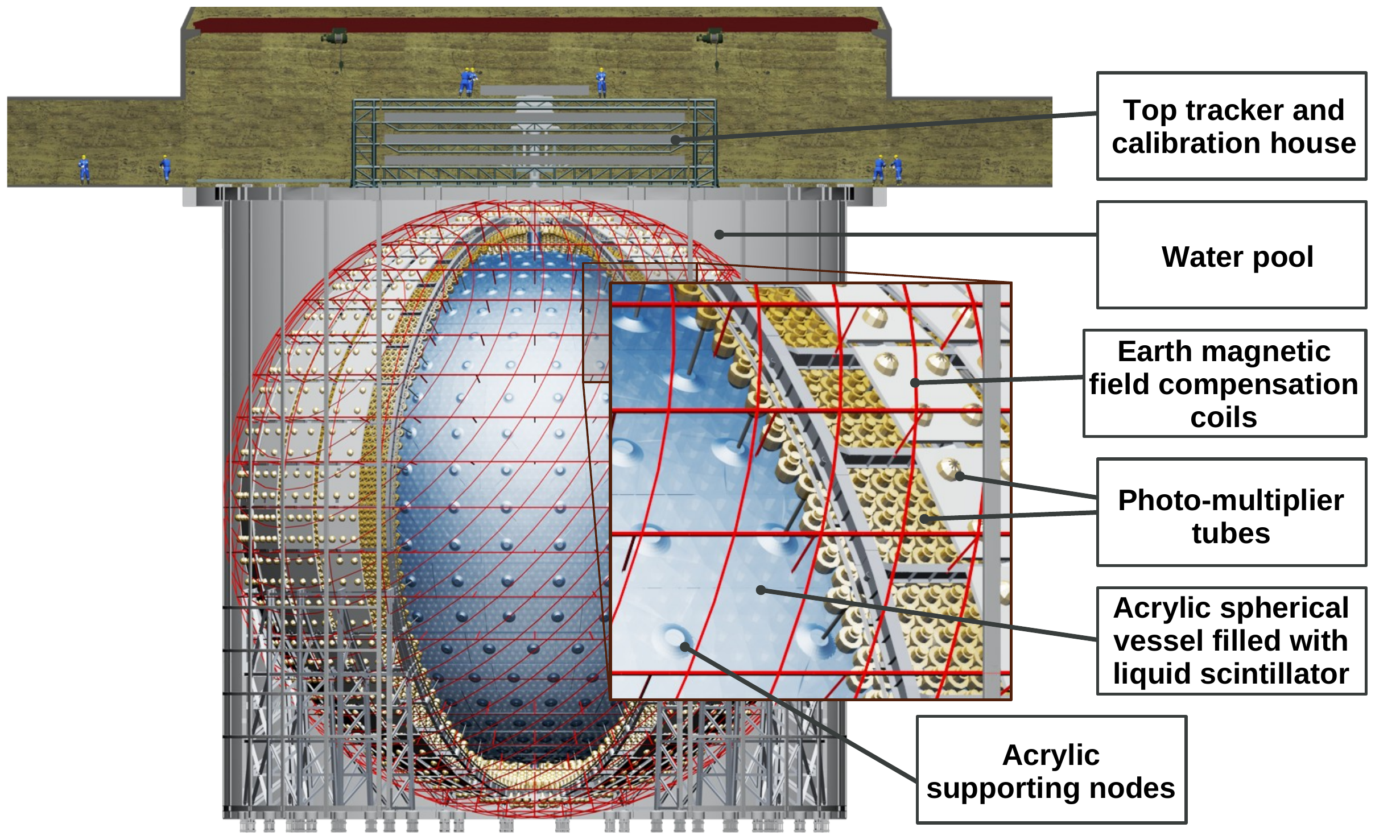}
	\caption{Schematic of the JUNO detector}
	\label{fig:detector}
\end{figure}

The JUNO central detector (CD) is composed of a spherical acrylic volume containing 20,000 tonnes of \textcolor{\mycolor}{the} liquid scintillator (LS), instrumented with 17,612 20-inch photomultiplier tubes (LPMT) and 25,600 3-inch photomultiplier tubes (SPMT) with photocathode coverage of 75\% and 3\%, respectively. The central detector is submerged in a water pool (WP) which is instrumented with 2,400 LPMTs providing detection of Cerenkov light from muons and shielding the LS from naturally occurring radioactivity of the surrounding rock. Precise muon track measurements are also provided by the top tracker (TT) detector mounted above the water pool. A schematic view of the detector is presented in Fig.~\ref{fig:detector} and further details are outlined in prior publications{\color{\mycolor}{~\cite{JUNO:2022hxd,JUNO:2015sjr}}}.%

Simulation software is an essential part of the JUNO experiment, providing a flexible and consistent interface to perform simulation studies across a broad physics program. The JUNO simulation software is based on the Geant4~\cite{GEANT4:2002zbu,Allison:2006ve,Allison:2016lfl} toolkit and the SNiPER framework (Software for Non-collider Physics Experiment)~\cite{Zou:2015ioy}, which is an experiment independent open-source project with sources available from github~\cite{sniper}. The software has been successfully used to perform massive Monte Carlo production at both local computing resources and distributed computing resources. The MC samples produced have been successfully used for the final selection and optimization of the detector design scheme and have been vital for the development of improved vertex, energy and track reconstruction algorithms. In addition they have enabled definition of the calibration strategy, provided evaluations of natural radioactivity background requirements, and facilitated development of online event classification algorithms, and assisted many physics studies~\cite{Li:2015cqa,JUNO:2020xtj,Li:2021oos,Qian:2021vnh,Li:2022tvg,Zhang:2018kag,Yang:2022din,JUNO:2021kxb}. In the past the primary focus of liquid scintillator neutrino experiments was the detection of low energy events, typically below tens of MeV. However for JUNO a much broader energy range is possible from tens of keV up to the level of GeV. The broad energy range and the scale of the JUNO detector with the world’s largest scintillator volume present data processing challenges across all areas. This paper addresses the simulation challenges and describes the detailed simulation software implementation in JUNO.%
%
%
\section{Requirements and Challenges of JUNO Simulation}%

The simulation software aims to achieve a precise reproduction of 
experimental data across a broad range of physics analyses in JUNO and 
is expected to have a software lifetime exceeding 30 years. 
Attaining this goal requires a longterm commitment to ongoing development
that strives to continually improve accuracy, reliability, usability, efficiency,
maintainability, and portability. Important simulation challenges 
for this ongoing work are highlighted below. 
\begin{itemize}
    \item%
The rich physics program in JUNO leads to the large number of event
generators that are essential for the simulation. These generators are
implemented with various programming languages and differing output formats.
A unified and flexible generator interface is essential to integrate the
diverse variety of generators within the JUNO simulation in a cost-effective manner. 
Various generation tools are also required, for example generating events 
in given volumes, materials or positions within the detector. 
In addition, the simulation software should also support the
dynamical deployment of calibration sources and the corresponding support structures in the simulation. 
The proposed solutions are discussed in section~\ref{sec:generator}.
    \item%
The simulation requires a large number, of input parameters for the construction
of detector geometries and various properties of the simulation including 
models of the LS, PMT and electronics. Some of these parameters are used by 
other offline data processing stages in addition to the simulation. 
Consistency of the parameters across the different stages must be guaranteed. 
During Monte Carlo tuning, it is also important to be able to vary parameters 
in a straightforward and flexible way. 
The parameter management and access strategy is described in section~\ref{sec:parameter} 
and section~\ref{sec:geometry}.
    \item%
Mixing simulation events from various sources allows an imitation of real data
to be created. However, due to the importance of time correlation and long-time
scales for some event types, it is challenging for mixing implementations 
to fit within memory budgets. For example, inverse beta decays (IBD) 
yield a positron which can form a prompt readout and a neutron which can form a 
delayed readout from neutron capture with \textcolor{\mycolor}{the} average time interval of about 200~$\mu$s.
Other decays of radioactive isotopes arising from muons or U/Th decay chains, 
can yield multiple products with time intervals of milliseconds or even longer. 
A further consideration is that mixing needs to be done prior to the electronics 
simulation (at hit level) in order to correctly treat multiple events falling 
within the readout window (pile-up). 
As loading all events relevant to long time intervals into memory 
at once would be prohibitively expensive a novel so-called {\color{\mycolor}{``pull''}} 
technique is developed to implement the mixing in a much more 
memory-efficient manner. This method is introduced in section~\ref{sec:elecsim}. 
   \item%
Monte Carlo truth information is critical for multiple purposes during
data processing and analysis, including algorithm development, understanding
detector response, estimating errors, and calculating efficiency in the
reconstruction and physics analysis. Providing straightforward access to 
truth information is extremely useful in many situations. 
However, after event mixing, a physical process such as IBD could be 
recorded into multiple readout events. Conversely, one readout event
may have contributions from multiple physical events. These 
complications make it non-trivial to provide full details of 
the relationship between truth information and the readout events.
The proposed solution is presented in section~\ref{sec:evt_corr}.  
    \item%
The simulation of high energy events within JUNO {\color{\mycolor}{is}} important 
to estimate backgrounds induced by cosmic muons and also for \textcolor{\mycolor}{the} studies
of atmospheric neutrinos and nucleon decays. 
High energy events can yield huge numbers of optical photons in the LS, 
for example, tens of millions of photons can be generated from a cosmic muon 
with a typical energy of 200~GeV crossing the CD. 
It is challenging to handle the optical photon simulation at this
scale within affordable computing resources.
We propose several methods to address this issue, including the fast 
parameterization simulation and the GPU-based full optical photon simulation. 
They are extensively discussed in section~\ref{sec:fast_sim}. 

\end{itemize}


\section{Simulation Software Architecture}%

The simulation software is one of {\color{\mycolor}{the}} key modules in the JUNO offline software~\cite{Huang:2017dkh}. 
It depends on the SNiPER framework and several other external libraries 
including Geant4, CLHEP~\cite{Lonnblad:1994np}, Boost~\cite{boost} and ROOT~\cite{Brun:1997pa}. 
The current versions used are Geant4 10.4.p02 and ROOT 6.24 with upgrades foreseen.

{\color{\mycolor}{SNiPER is designed and developed with Object-Oriented technology and bi-language of C++ and Python. SNiPER has many innovations in the management of correlated events by introducing an event buffer mechanism, multi-task processing controlling, and fewer dependencies on third-party software and tools. SNiPER also reserves the interfaces to the implementation of multi-threading computing.}} 
All modules in the offline software are defined as dynamically loadable elements (DLE) in SNiPER, 
which can be loaded and executed dynamically at run time. The modules to load 
are configured with Python using interfaces provided by SNiPER.  
The DLEs provide flexible workflow control and memory management. 

SNiPER roles distinguish the modules into algorithms, services, tasks, and tools. 
These roles are similar to those of the Gaudi framework~\cite{gaudi}. 
Algorithms are called once per event during the event loop, they create 
data objects and process data objects in the memory. 
Services \textcolor{\mycolor}{provide} common functionalities, like accessing detector geometry
information, and provides access to other parts of the framework. \textcolor{\mycolor}{A} task is a
lightweight application manager that orchestrates DLEs and controls the event
loop. A SNiPER application always starts from a task. The tool is a lightweight
routine that enables sharing of a specific functionality between algorithms.
In addition to module control, SNiPER also provides in-memory data buffer management, 
job configuration with variable customization, logging information with various output levels,
and incident handlers following the Gaudi approach. 

\begin{figure}[htb]\centering
	\includegraphics[width=85mm]{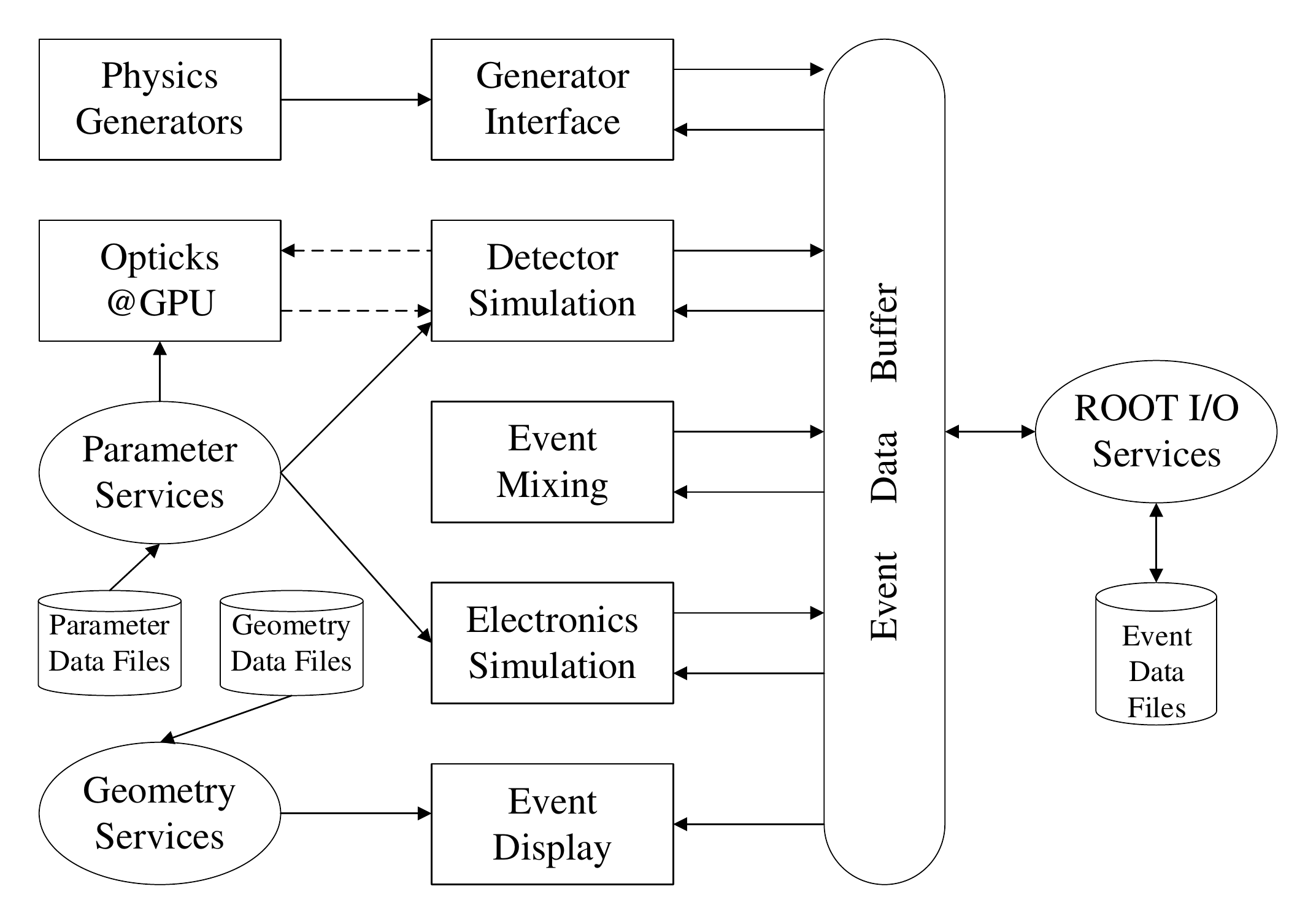}
	\caption{Architecture of the JUNO simulation software}
	\label{fig:workflow}
\end{figure}

The architecture of simulation software is shown in Figure~\ref{fig:workflow}.
It consists of four SNiPER algorithms of generator interface, detector
simulation, event mixing and electronics simulation, a few SNiPER services for
parameter access, and data I/O management. And it also includes several components that are not
strongly coupled or independent with the SNiPER framework, such as physics
generators, event display, and Opticks~\cite{Blyth:2021gam}. The algorithms are
independent \textcolor{\mycolor}{of} each other. They first read the input data from the event data
buffer if necessary, then write the output data back to the buffer after the
processing. The buffer data objects are persisted via the ROOT I/O service~\cite{Li:2017zku}. 
The parameter services and geometry services~\cite{Li:2018fny,Zhang:2020jkg} provide requested
parameters, and geometry information to other components in the simulation software. 
Opticks is an open-source package providing GPU-accelerated optical photon simulation 
that is integrated with the detector simulation. 
Two event display applications~\cite{You:2017zfr,Zhu:2018mzu} are
also developed for debugging, validation, and public outreach.
Further details on these components are presented in the following sections.%

%
%


\section{Physics Generators}
\label{sec:generator}

\begin{figure}[htb]\centering
	\includegraphics[width=65mm]{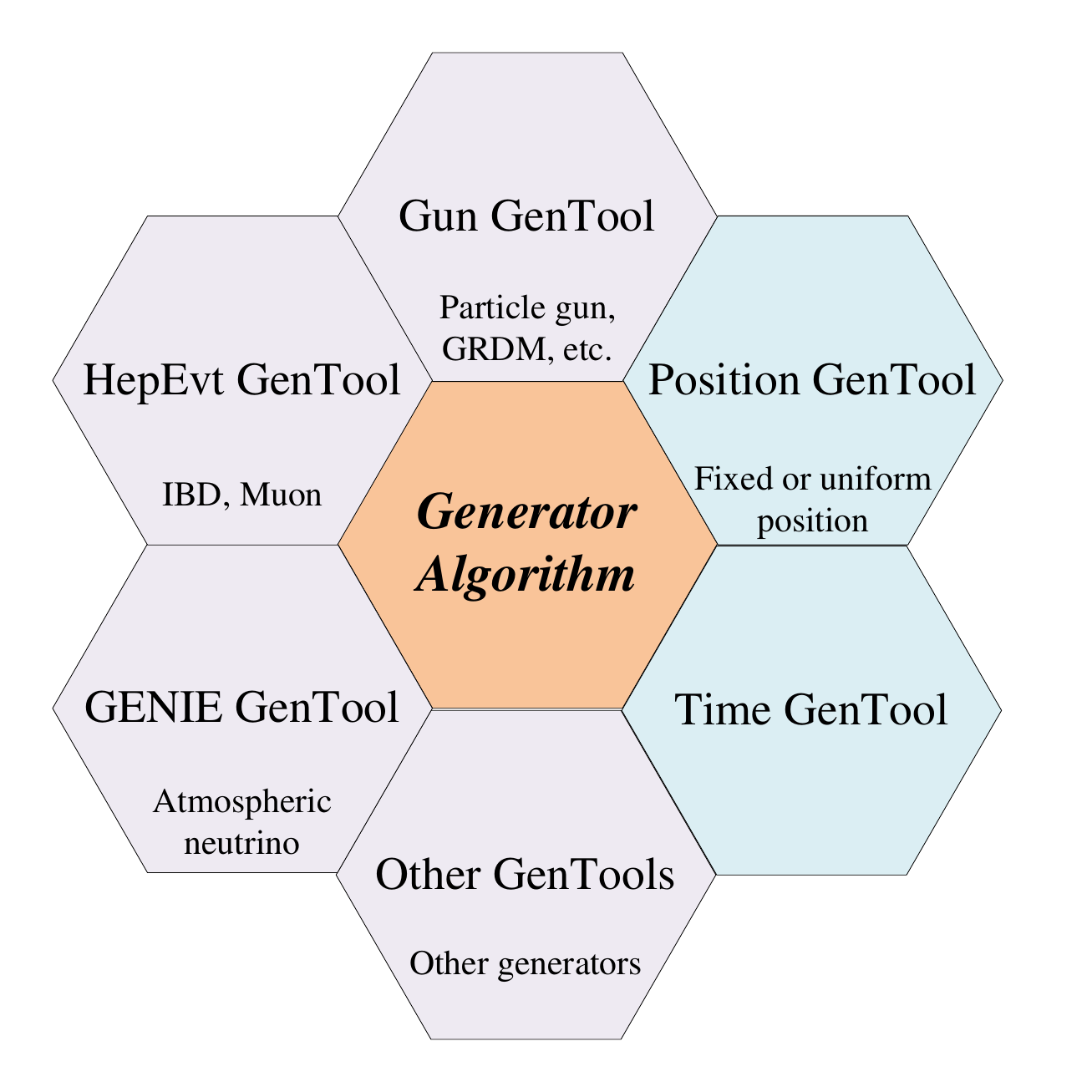}
	\caption{Overview of physics generator software}
	\label{fig:generator}
\end{figure}

A variety of physics generators has been developed or integrated with the JUNO offline software system to meet the needs of a broad physics program. The physics generators include generators for cosmic muons, reactor neutrinos, atmospheric neutrinos, solar neutrinos, geo-neutrinos, supernova burst neutrinos, {\color{\mycolor}{DSNB }}(Diffuse supernova neutrino background), natural radioactivities, and calibration sources.
Most are implemented as standalone packages with 
differing output formats. A modular physics generator interface 
is developed, as a SNiPER algorithm, to unify the generation of primary particles 
from various generators. The software design is illustrated in Figure~\ref{fig:generator}.%
In the design, the generator algorithm manages a sequence of SNiPER tools. 
The HepMC event record~\cite{Dobbs:2001ck} is used as the intermediate data object 
between the generator algorithm and the generator tools. 
A \texttt{GenEvent} object in HepMC represents a physical event,
containing any number of vertex objects named \texttt{GenVertex}; each
\texttt{GenVertex} represents an interaction point with the
generated final state particles represented by \texttt{GenParticle} objects. 
The generator algorithm invokes the list of generator tools in sequence
to build the HepMC objects. The HepMC objects are subsequently converted 
into Geant4 primary vertex objects for use by the simulation. Each
generator tool updates the HepMC objects based on the output
information of the corresponding physics generator.%

This design allows flexible integration of any physics generator of interest, 
via two principal methods:
\begin{itemize}
    \item%
A standalone physics generator produces output data files in disks or buffers first, then the corresponding generator tool reads them and converts the format provided into HepMC objects.
    \item%
A physics generator is re-implemented as a generator tool which 
directly produces HepMC objects.%

\end{itemize}
The particle gun \texttt{GenTool} generates particles with dedicated particle
types, momenta and positions. 
The particle type can be specified with a PDG integer code, a particle name string, 
or an input file path. Particle name strings are parsed into PDG codes using ROOT \texttt{TDatabasePDG}. 
Using an input file provides an interface to generation of radioactive isotope decays 
using the Geant4 Radioactivity Decay Module (GRDM).
The input file format specifies multiple radioactive isotopes with each line 
providing PDG code, mass, ratio and ranges of atomic number (Z) and mass number (A). 
During the particle generation the particle type is randomly sampled according to the ratio
of each specified isotope. Limiting the ranges of Z/A for each radioactive isotope, 
allows a subset of the decay chain to be simulated.

The HepEvt \texttt{GenTool} converts events from HepEvt format to the HepMC
format. A physics generator can be executed in advance of the simulation job
and generate an output file with the HepEvt format, then the HepEvt
\texttt{GenTool} reads events from the generated output file and performs the
conversion to HepMC. Alternatively, the physics generator can be executed with \texttt{popen} 
using a Unix pipeline to avoid the need for intermediate files.%

The GENIE \texttt{GenTool} uses another generator integration strategy, where 
the GENIE~\cite{Andreopoulos:2009rq} libraries are linked to the simulation software 
and its functions are invoked from the GENIE \texttt{GenTool}. 
The GENIE output in \texttt{GHEP} data format is converted to the HepMC data format. 
The parameters needed to configure GENIE are set as properties of the
GENIE \texttt{GenTool} that can be set from \textcolor{\mycolor}{the user's} Python scripts.  

The position \texttt{GenTool} facilitates event generation
either at fixed positions or uniformly distributed throughout a volume.  
Positions can be specified within named detector volumes or materials.  
Separation of this access to detector geometry information prevents 
duplication within other gentools. 
During \textcolor{\mycolor}{the} initialization of this tool, a recursive traversal of the Geant4 geometry 
is performed which caches transformation matrices for the volumes. 
When generating a position, a volume is selected according to the user
specified volume name, then a bounding box is calculated by the Geant4 solid.
Positions inside the bounding box are generated randomly and only the positions
inside the solid are selected. Then the relevant cached transform is applied 
to obtain global positions from the generated local positions.  
In addition, global position cuts may be applied to limit the generated positions, 
such as restricting to a shell within a radius cut, or generating positions
within a layer using cuts along different axes. 
Subsequently, the Geant4 \texttt{G4Navigator} is used to access the material 
at this global position, allowing comparison with the user-specified material.
The sampling is repeated until the user's requirements are fulfilled.%

The time \texttt{GenTool} is used to control the timestamp of interaction
vertices for each event and maintain the time correlation among vertices. 
In most cases, the timestamp of an event starts as zero in the detector simulation 
and is re-assigned in the electronics simulation.
However, there are some exceptions, such as the supernova burst events, in which the timestamp is defined by the generator and used directly in the subsequent electronics simulation.

One challenge as mentioned previously is the handling of the 
dynamic deployment of the calibration sour-ces. 
In order to generate particles from these calibration units 
from various positions without requiring multiple geometries,
the units are implemented as detector elements, which can be
placed into the containing detector element dynamically. 
The position \texttt{GenTool} traverses the Geant4 geometry to 
find calibration units by name and material. 
Subsequently, positions are sampled uniformly within the local volume 
and then transformed into global positions.


\section{Detector Simulation}
\subsection{Integration of Geant4 with JUNO simulation}
\begin{figure}[htb]\centering
	\includegraphics[width=85mm]{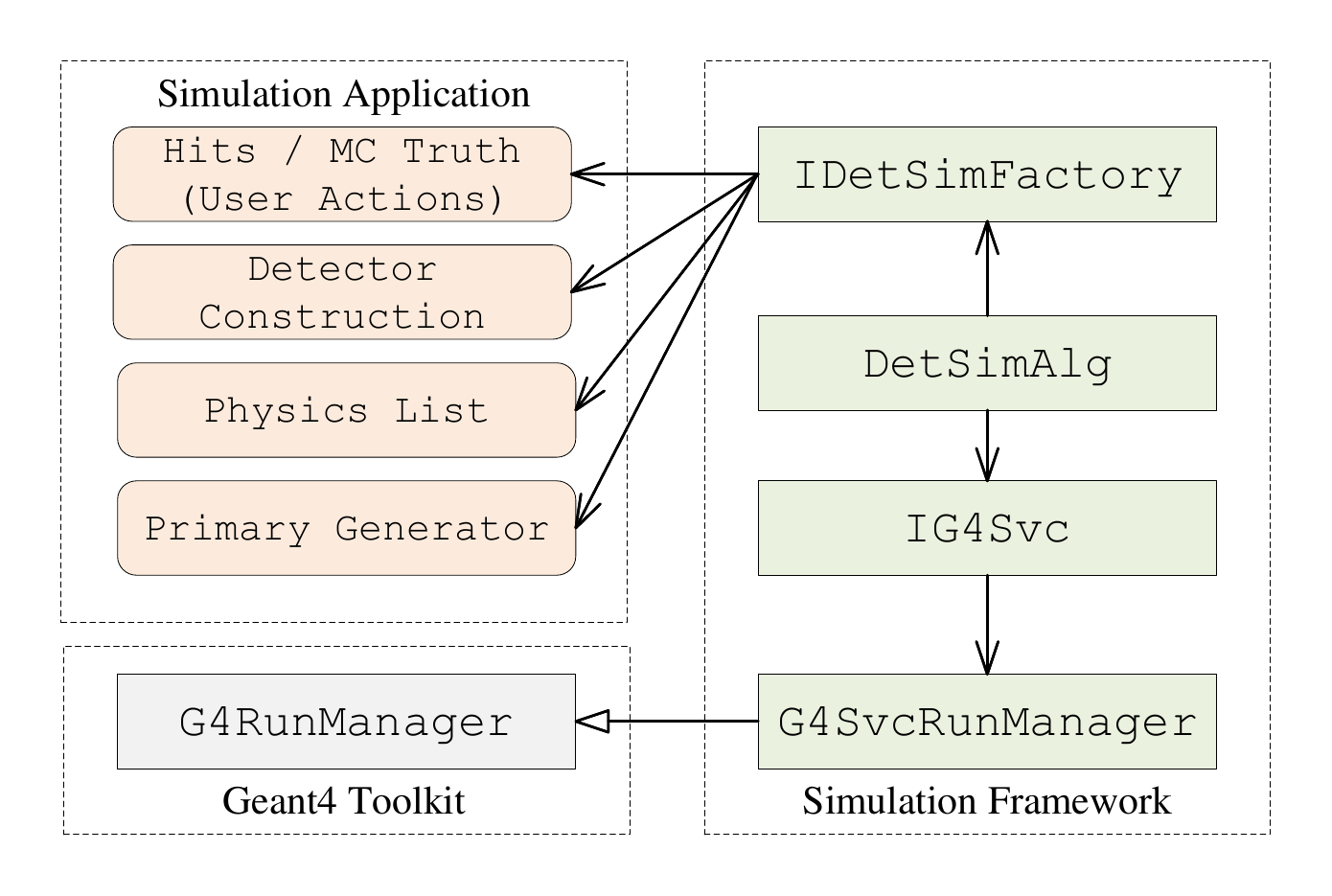}
	\caption{Integration of Geant4 toolkit with JUNO simulation}
	\label{fig:integration}
\end{figure}

{\color{\mycolor}{Both SNiPER and Geant4 have their own event loop control. A general simulation framework has been developed to adapt Geant4's event processing workflow into SNiPER. The integration strategy of Geant4 and SNiPER}} is illustrated in Figure~\ref{fig:integration}. 
The simulation framework comprises algorithms and services that adapt the 
original Geant4 workflow, acting as a bridge to provide flexible 
control of the underlying simulation. 
This approach allows most of the code from a standalone use of Geant4 
to be used unchanged within the integration.%
%

As shown on the right of Figure~\ref{fig:integration}, the simulation framework consists of four parts: an algorithm named \texttt{DetSimAlg}, which is the entrance to detector simulation; an implementation of a customized run manager named \texttt{G4SvcRunManager}, which is inherited from the \texttt{G4RunManager} in Geant4 toolkit to allow simulation of a single event; a service named \texttt{IG4Svc}, which controls the run manager; a service named \texttt{IDetSimFactory}, which is a factory to create instances of detector construction, physics list, primary generator and user actions. The simulation workflow is fully controlled with these components. 
At \textcolor{\mycolor}{the} initialization stage, \texttt{DetSimAlg} invokes instances of \texttt{IG4Svc} and \texttt{IDetSimFactory} to create the customized run manager which allows the Geant4 kernel to be initialized. At \textcolor{\mycolor}{the} execution of the event loop, \texttt{DetSimAlg} invokes the run manager to simulate an event. In addition to the normal event loop, starting the Geant4 UI manager for visualization is also supported when users specify the \texttt{--vis} command line option.

As shown on the left of Figure~\ref{fig:integration}, a Geant4 simulation application consists of several parts: a detector construction to initialize the geometry; a physics list to initialize the particles and physics processes; a primary generator action to generate particles; user actions to handle detector response in sensitive detectors and other MC truth information. Some aspects of the simulation application can be customized using several SNiPER tools. %
For detector construction, a detector element named \texttt{IDetElement} is used to represent a high-level detector component and manage the corresponding logical volumes in Geant4. The placements of one detector element into another detector element are handled with \texttt{IDetElementPos}, which returns the positions. Users can enable, disable, or modify the detector geometry easily, fulfilling requirements such as placing a calibration unit at a fixed position, changing the arrangement of PMTs, or disabling some detector {\color{\mycolor}{components. }} 
Modular physics constructors, provided in Geant4, are used to build the physics list in the JUNO simulation. Each modular physics constructor is managed by a SNiPER tool to support the configuration of parameters used in physics processes and models. This provides a flexible way for users to choose the proper physics models and parameters based on their equirements.

Geant4 user actions such as \texttt{UserSteppingAction} provide single entry points to access Monte Carlo truth information. However different analyses require access to different aspects of the truth information, presenting an organizational problem. This issue is solved by defining an \texttt{IAnalysisElement} interface which allows a user-configurable list of analysis elements to be composed where all have access to the MC truth information. The selection of truth information is stored within the Geant4 user track information, which is retrieved at the end of the event and stored \textcolor{\mycolor}{in} output files. 

\subsection{Parameter management}
\label{sec:parameter}

The simulation software requires a large number of parameters to define the geometry of the detector and the optical properties of \textcolor{\mycolor}{the} materials and surfaces, physics processes, and characteristics of the electronics. Dedicated parameter access services have been implemented to provide consistent access to parameters from various applications and to facilitate flexible parameter variation during Monte Carlo tuning.

Examples of the optical properties of materials provided by the parameter service are refractive indices, absorption and scattering lengths, scintillator light yield, Birks constants, emission spectra, and time constants. Many of the parameters are energy dependent. The parameter service uses keys such as \texttt{Material.LS.RINDEX} to control the parameters to retrieve. Command line configuration allows default parameter values to be overridden. 

Measurements of PMT parameters including quantum efficiency, collection efficiency, time spread, dark counting rate, gain, and resolution are stored in ROOT files with entries for every PMT. A dedicated PMT parameter service is implemented to provide access to these values to both the detector and electronics simulations using PMT identifier integers to select the PMT. 

\begin{figure}[htb]\centering
	\includegraphics[width=85mm]{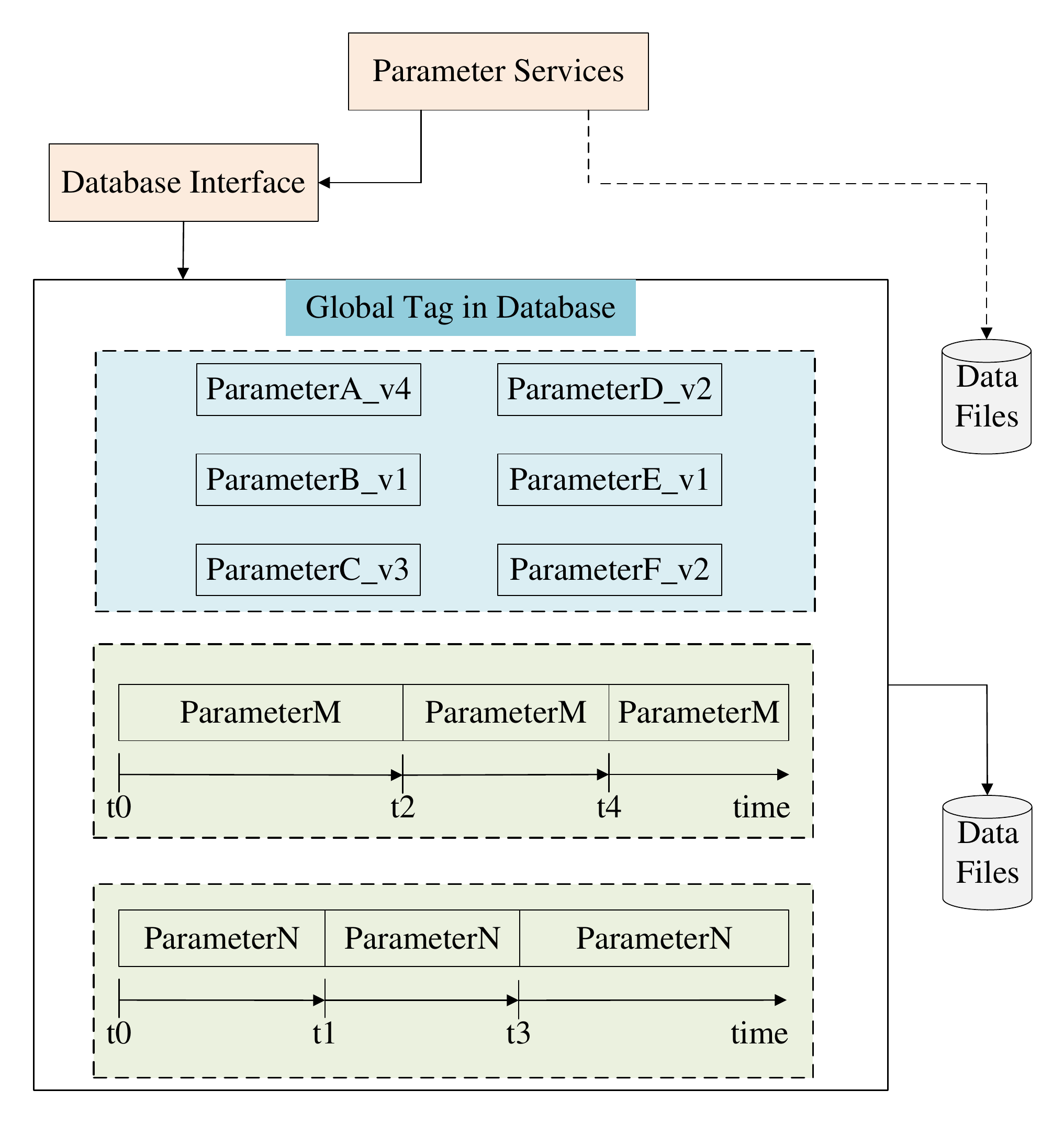}
	\caption{Parameter management in simulation software. The versions of static parameters (in blue) and the time intervals of condition parameters (in green) are saved as Global Tag in database.}
	\label{fig:parameter}
\end{figure}

The parameters used in \textcolor{\mycolor}{the} simulation can be divided into static parameters which vary only infrequently with software updates and conditions data which change during data taking and require periodic calibrations with associated time ranges. Figure~\ref{fig:parameter} illustrates how static parameters (in blue) and condition parameters (in green) with associated time intervals are grouped together with a \texttt{GlobalTag} that labels coherent parameter sets. For official data production, each offline software version will have one \texttt{GlobalTag} identifying the parameter set. The parameter services use a database interface to retrieve metadata that enables the parameters associated with the \texttt{GlobalTag} to be retrieved from the production file system. For MC tuning or user testing, it is possible to override the parameter files used.

{\color{\mycolor}{The detector simulation, electronics simulation, and reconstruction algorithms handle the conditions data and parameter management in the same way. }}

\subsection{Geometry construction}
\label{sec:geometry}

\begin{figure}[htb]\centering
	\includegraphics[width=85mm]{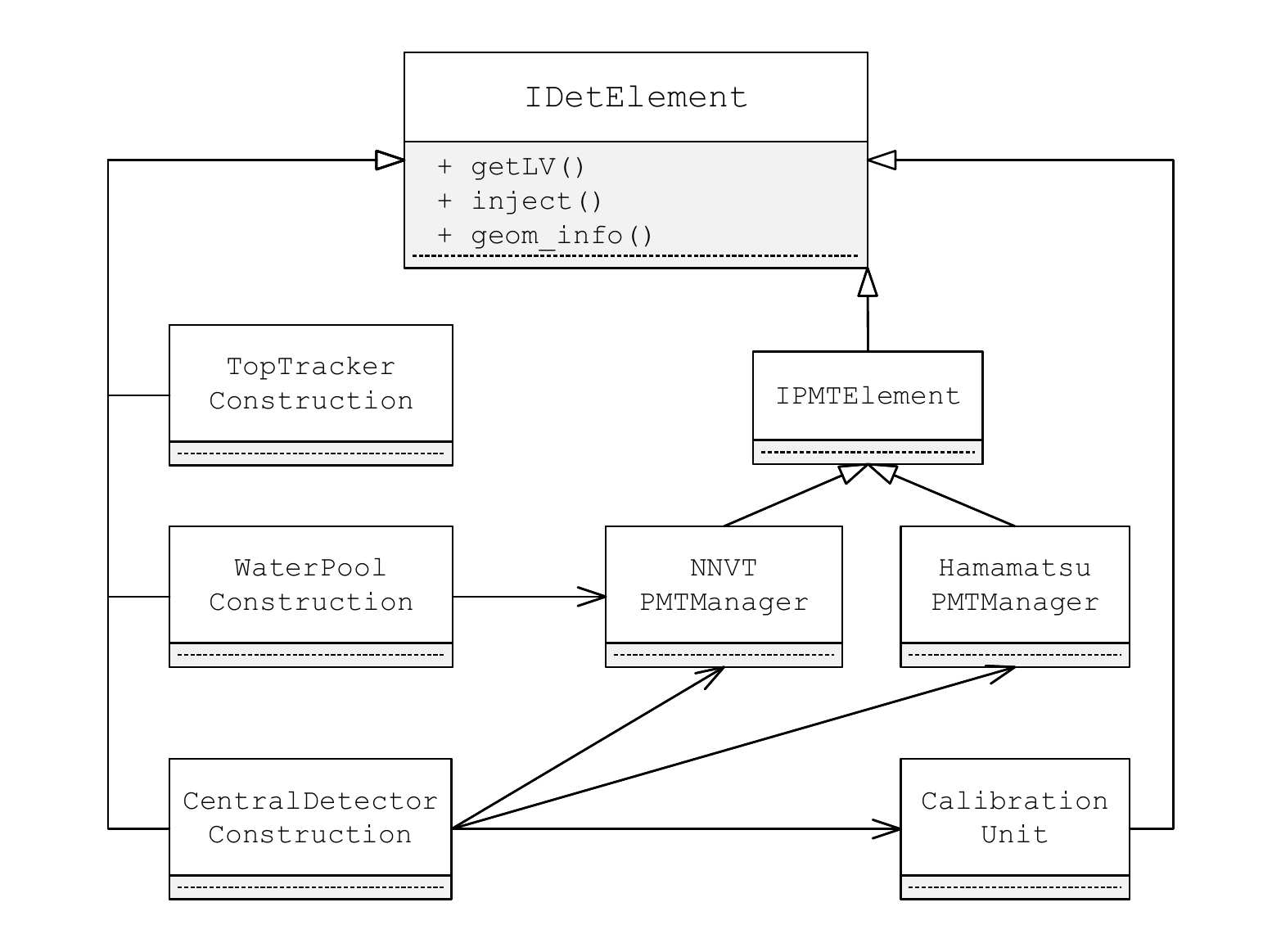}
	\caption{Design and use of \texttt{IDetElement} detector elements}
	\label{fig:DetectorElement}
\end{figure}

Flexible geometry management is particularly beneficial  
during detector design and optimization studies. 
The primary technique used to maintain a flexible geometry 
is to factorize the geometry into groups of volumes using the 
\texttt{IDetElement} base class and to mediate the composition 
of those detector elements using \texttt{inject} methods 
designed to retain {\color{\mycolor}{independence}} between the detector elements.  
Figure~\ref{fig:DetectorElement} illustrates the usage of these detector elements. For example, the central detector is a detector element which directly contains only the liquid scintillator, its container, and the buffer material. The PMTs are not directly placed within the central detector, but their placement is mediated using the \texttt{IDetElement} instead. This simplifies interchanging different options for the central detector while keeping the arrangement of PMTs unchanged. This {\color{\mycolor}{independence}} of the target geometry from the PMTs also proved useful for investigations of different types of PMTs.
%

Detector components and materials can be defined directly in Geant4 code or indirectly via parsing GDML (Geometry Description Markup Language) files or text files. Simulation of the calibration geometry elements, such as movable radioactive sources, \textcolor{\mycolor}{uses} a separate runtime geometry placement system.

\begin{figure}[htb]\centering
	\includegraphics[width=65mm]{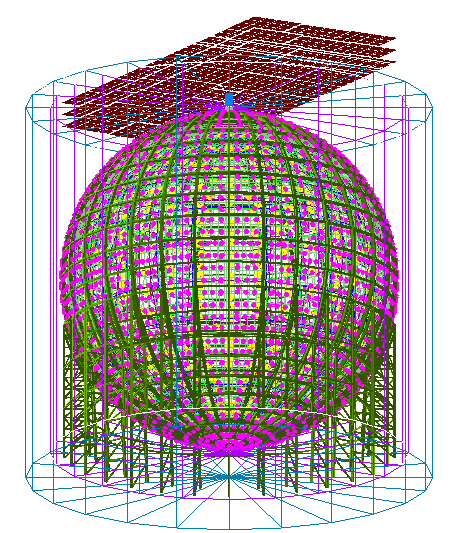}
	\caption{JUNO detector in simulation software. 3 layers of top tracker (red), 2400 WP PMTs (pink), 590 acrylic nodes (yellow), stainless steel truss (green) and chimney (blue)}
	\label{fig:JunoSimDetector}
\end{figure}

The geometries of CD, WP, TT, calibration units, and other supporting structures in the JUNO detector have all been implemented in the simulation software according to the final design. As shown in Figure~\ref{fig:JunoSimDetector}, the simulated JUNO detector includes 3 layers of top tracker (red), 2400 WP PMTs (pink), 590 acrylic nodes (yellow), stainless steel truss (green), and chimney (blue). The 17612 20-inch PMTs and 25600 3-inch PMTs in {\color{\mycolor}{the}} CD, calibration anchors inside and outside the acrylic sphere, and guide tube have also been implemented in the simulation software but not displayed in this figure in order to display other parts more clearly. 

\begin{figure}[htb]\centering
	\includegraphics[width=85mm]{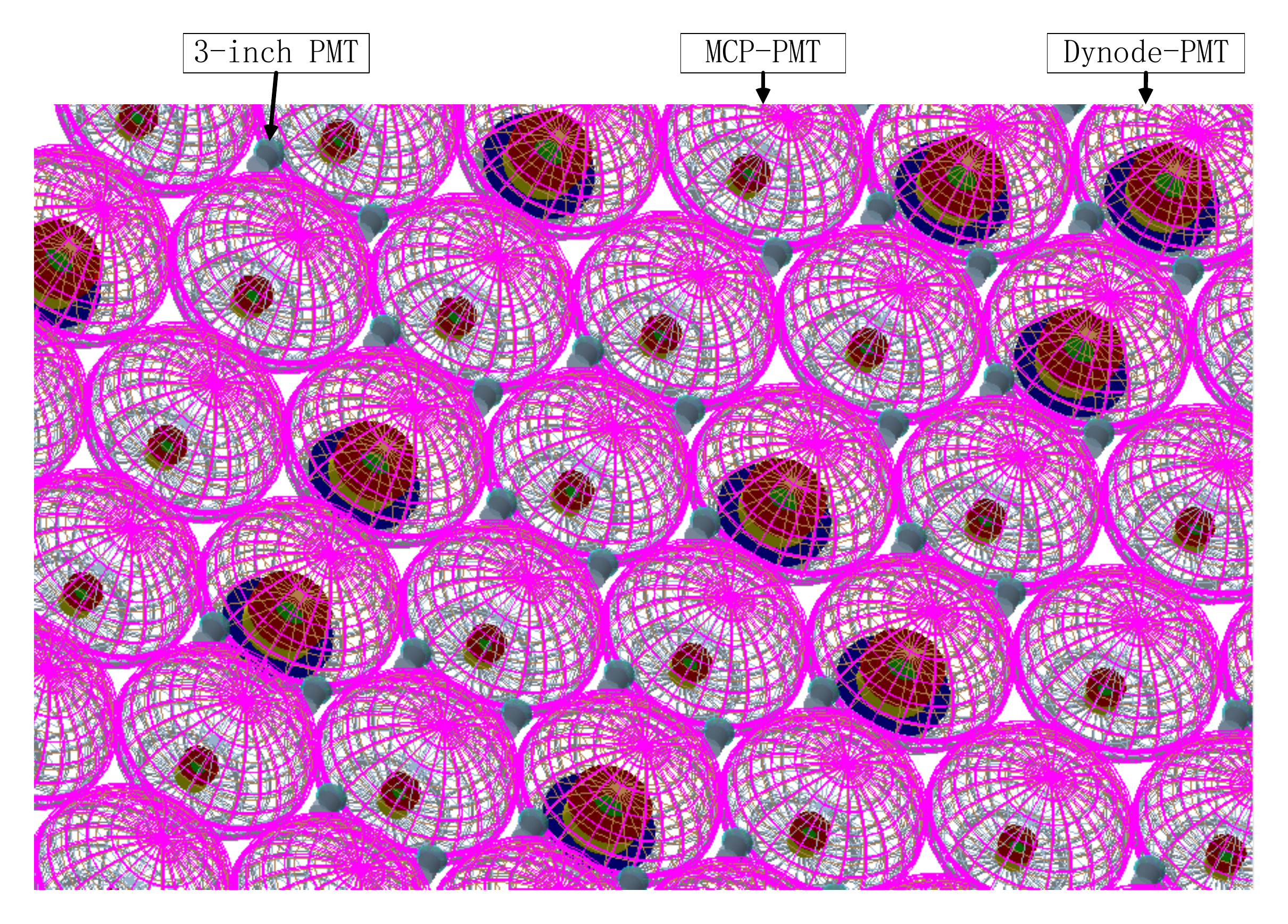}
	\caption{{\color{\mycolor}{Two types of 20-inch PMTs in CD, including dynode-PMT and MCP-PMT indicated with labels. The 20-inch dynode-PMTs have much larger electrodes inside the PMTs, compared with the MCP structures in the MCP-PMTs. The 3-inch PMTs are represented as the grey color and located in the gaps among 20-inch PMTs, in which only about 85$\%$ of gaps are filled with 3-inch PMTs. The acrylic protection cover is shown in mesh mode with a magenta color.}} }
	\label{fig:PMTmask}
\end{figure}

As shown in Figure~\ref{fig:PMTmask}, in the central detector, two types of 20-inch PMTs are implemented in the detector simulation, including dynode-PMT produced by Hamamatsu and MCP-PMT manufactured by NNVT. Besides the PMT glass and inner vacuum, the PMT inner components, acrylic protection cover at the front, and steel protection cover at the back are also implemented. PMT inner components include a cylindrical tube at the bottom, a focusing electrode on the top, and a dynode (MCP) in the center of the focusing electrode. 
It is found that containing the PMT and protective cover within a virtual volume speeds up the Geant4 geometry initialization, presumably due to the complicated structure and large number of PMTs.
The actual installation mounts PMTs into apertures through an optical mask used for optical isolation between the outer water pool and CD. To implement the optical mask geometry and account for the reflectivity of \textcolor{\mycolor}{the} mask in the simulation whilst avoiding expensive boolean geometry the tail of the 20-inch PMT is shortened to {\color{\mycolor}{avoid}} overlap between the PMTs and optical mask. As optical effects arising from the tail of the PMT on the other side of the mask are expected to be negligible the PMT tails are not simulated. 
%
%
The detector geometry is defined using the Geant4 geometry model, while reconstruction, event display, and analysis packages are heavily dependent on ROOT. To provide a consistent detector description for different applications the GDML is chosen as the basis of the geometry management system~\cite{Li:2018fny}.
Detector simulation reads geometry parameters from text files that are used to construct the geometry using the Geant4 C++ geometry model API. The complete geometry is exported using the Geant4 GDML writer into GDML files that serve as input to the geometry service using ROOT GDML import functionality. The ROOT geometry objects resulting from the import are used by the geometry service to construct mappings between detector identifiers and corresponding objects.
The geometry service provides a detector information interface that is used by other applications including reconstruction, event display, and analysis. The consistency of \textcolor{\mycolor}{the} detector description relies on Geant4 GDML export and ROOT GDML import functionality.

\subsection{Physics processes}
Geant4 provides a set of physics constructors that are combined to build
reference physics lists. Physics processes for JUNO detector simulation are
created with the Geant4 physics constructors shown in Table~\ref{table:physics}.
{\color{\mycolor}{A}} Positronium process has been registered into the Livermore physics models 
\texttt{G4EmLivermorePhysics}.  A modified neutron capture process has been registered in \texttt{G4HadronPhysicsQGSP\_BERT\_HP}.

The Geant4 Radioactivity Decay Module (GRDM) has been used for JUNO background evaluation.
{\color{\mycolor}{The correct treatment of $^{9}$Li and $^{8}$He decays is of fundamental importance in the study of cosmogenic backgrounds for JUNO. In GEANT4, the cosmogenic nuclei undergo beta decay of $^{9}$Li and $^{8}$He reaching the correct state. 
However, for the de-excitation of the $^{9}$Be and $^{8}$Li nuclei, the correct decay chains including alphas and neutrons are not accounted for. 
Thus the GRDM and the related data files have been modified to correctly handle $^{9}$Li and $^{8}$He decays, }}as described in~\cite{Jollet:2019syr}. Such a
modification is now included in Geant4 since release 10.6.

For Ion processes, \texttt{G4IonPhysics} or \texttt{G4IonPhysicsPHP} can be chosen by users,
depending on the particles to be simulated and the memory resources available. {\color{\mycolor}{G4IonPhy-sicsPHP provides more reasonable simulation for the light nuclei at the MeV energy region, in which the G4TENDL data set is used. The TENDL cross-section table was substituted with that calculated by TALYS. ~\cite{Koning:2019qbo}}}
\begin{table}
\centering
\caption{Geant4 Physics constructors used by JUNO simulation}
\label{table:physics}
\setlength{\tabcolsep}{2pt}
\begin{tabular}{|l|c|}
\hline
Physics Constructors &  Status \\
\hline
\texttt{G4EmLivermorePhysics} & Customized \\
\texttt{G4EmExtraPhysics}  &  Unchanged \\
\texttt{G4DecayPhysics}  &   Unchanged\\
\texttt{G4RadioactiveDecayPhysics} &   Customized \\
\texttt{G4HadronPhysicsQGSP\_BERT\_HP}   &  Customized \\
\texttt{G4StoppingPhysics}  &  Unchanged \\
\texttt{G4IonPhysics} or \texttt{G4IonPhysicsPHP}  &    Unchanged \\
\texttt{G4OpticalPhysics}   &  Customized \\
\hline
\end{tabular}
\end{table}

Geant4 processes that model optical photon propagation are
\texttt{G4OpBoundaryProcess}, \texttt{G4OpRayleigh} and \texttt{G4OpAbsorption}, and 
are used by JUNO simulation. These processes use material properties 
such as refractive indices, absorption and scattering lengths and optical surface properties 
such as reflectivity which are set according to measurements. 
The Geant4 processes that produce optical photons are \texttt{G4Scintillation} and \texttt{G4Cerenkov}. 
They were both modified to meet the needs of JUNO simulation. 
\texttt{G4Cerenkov} assumes a refractive index that monotonically increases with photon energy.
As this is not the case for the JUNO LS refractive index the \texttt{G4Cerenkov} process was 
customized to handle more general refractive index characteristics.
The \texttt{G4Scintillation} process was customized in several ways:
\begin{enumerate}
\item Different Birks constants were set for different particles.
\item Particle specific emission time constants and exponential decay
components were used for gammas, electrons, positrons,
alphas, neutrons and protons.
\item  Photon reemission was implemented for the liquid scintillator, as optical 
photons arising from both scintillation and Cerenkov processes 
can be absorbed in the liquid scintillator and then re-emitted
with a different wavelength. The reemission probability depends on wavelength.
\end{enumerate}
%
Several physics constructors have been defined as SNiPER Tools 
to facilitate flexible control and customization. For example 
users can easily switch on/off the scintillation or Cerenkov
processes or customize properties of each process.

\subsection{JUNO PMT optical model}
The JUNO PMT optical model~\cite{Wang:2022tij} 
accounts for both light interactions with the PMT window
and optical processes inside the PMTs. 
The PMT optical model describes the angular 
and spectral response of the photon detection efficiency.
A newly developed package uses the transfer matrix method (TMM) 
from optics to account for interference effects within the thin layers 
of anti-reflection coating and photocathode between the PMT 
glass and interior vacuum.  Using inputs of layer thicknesses 
and complex refractive indices the TMM calculation yields 
reflectance, transmittance, and absorption of the multifilm stack of layers. 
 
The JUNO PMT optical model is implemented as a Geant4 fast simulation model
to integrate with the rest of the simulation.
The fast simulation model is triggered only for optical photons 
within an envelope inside the PMT glass.  
\subsection{Hit persistency}

After the simulation of PMT response, hits are created and
registered into Geant4 hit collections. At the end of each event, 
the Geant4 hits are converted into the JUNO event data model (EDM) SimHits
and saved into ROOT files. The EDM and ROOT I/O are described in section~\ref{sec:evt_corr}.

For low energy events with energy deposit of 1~MeV the number of hits is 
less than 2000, with acceptable memory consumption 
of 800 MB with G4IonPhysics or 1.6 GB more with G4IonPhysicsPHP. 
The more memory consumption with the higher precision 
model arises from additional datasets that are loaded into memory. 

\begin{figure}[htb]\centering
 	\includegraphics[width=55mm]{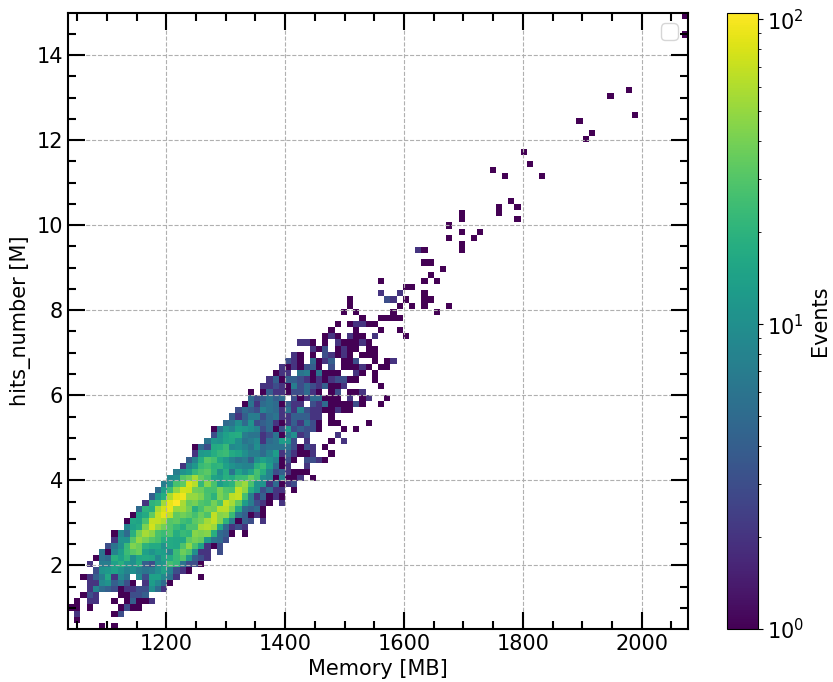}
 	\caption{Memory consumption versus number of hits in muon simulation}
 	\label{fig:MuonMemory}
\end{figure}

High energy events, such as muons crossing the CD, can yield millions of simulated hits 
with memory consumption of up to 20~GB which poses operational difficulties. 
Several strategies have been developed to reduce this memory consumption. 
\begin{enumerate}
\item Use smaller summary hits for high energy events that are about 20\% the size of full hits and which \textcolor{\mycolor}{contain} only vital information such as arrival time and the number of photo-electrons. 
\item Merge hits that arrive within a configurable time window. Using 1~ns
to match the 1~GHz waveform sampling of the readout electronics reduces 
the number of hits to about 30\% of the original for a muon of typical 
energy 215~GeV crossing the central detector. 
\item Split processing of events that exceed configurable memory limits into sub-events.
The sub-events are merged within the electronics simulation with sequential reads. 
ROOT writing is found to consume extra memory whilst simultaneously holding  
uncompressed and compressed buffers. 
\end{enumerate}
Without split processing, the maximum memory consumption for muon events can still exceed 6~GB. With splitting this maximum is reduced to about 2~GB for 14 million hits, when using G4IonPhysics, as shown in Figure~\ref{fig:MuonMemory}. This level of memory consumption fits within available computing resources allowing large scale production of simulation samples. 
%
%
%
%


\section{Event Mixing and Electronics Simulation (ElecSim)}
\label{sec:elecsim}
%
\subsection{Electronics simulation software package}
Package \texttt{ElecSim} is used to model the PMT response to hits from the Geant4-based DetSim package and also to model the readout electronics of CD, WP and TT using an implementation based on SNiPER managed dynamically loadable elements (DLE).  
The package incorporates an event mixing implementation \textcolor{\mycolor}{that} combines different event types to create simulated MC data that \textcolor{\mycolor}{mimics} real experimental data. Event mixing uses a {\color{\mycolor}{``pull''}} based workflow (see \ref{sec:pullmode}) using SNiPER incident triggers that stays within reasonable memory budgets. 
ElecSim outputs become inputs to the online event classification algorithms (OEC) used for event tagging and may be saved to \textcolor{\mycolor}{files} using EDM formats and ROOT I/O services.  
Detailed descriptions of the JUNO PMT and electronics 
systems are available in prior publications{\color{\mycolor}{~\cite{JUNO:2022hxd,JUNO:2022hlz,Bellato:2020lio,Fang:2019lej}}}.

\begin{figure}[htb]\centering
        \includegraphics[width=85mm]{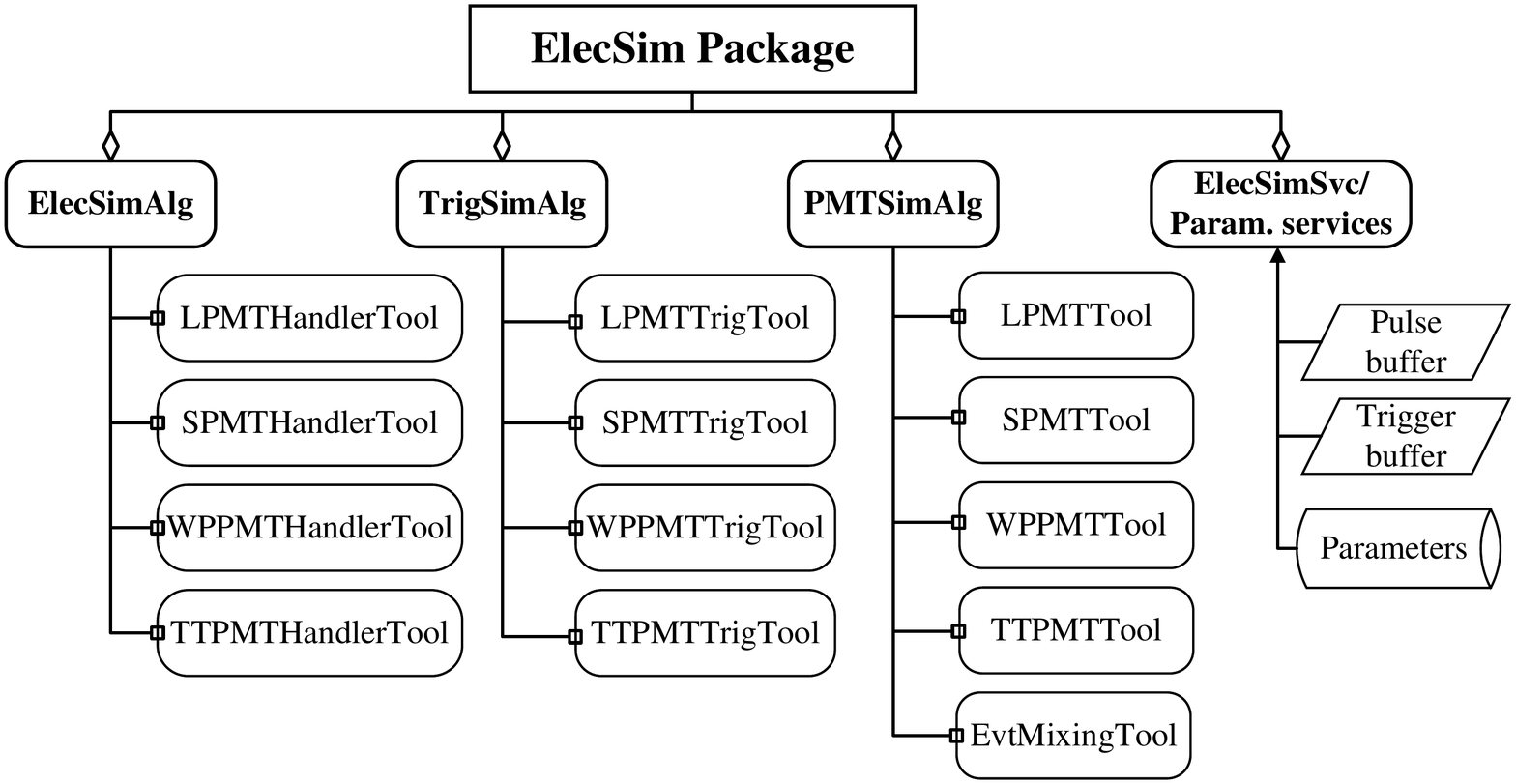}
        \caption{Structure of the electronics simulation software.}
        \label{fig:ElecSimSoftware}
\end{figure}

The components of the package are illustrated in Figure~\ref{fig:ElecSimSoftware},
consisting of several services and three algorithms \texttt{PMTSimAlg},
\texttt{ElecSimAlg} and \texttt{TrigSimAlg}. 
The services manage the temporary Pulse and Trigger buffers which provide communication 
between the algorithms and also provide parameter access. 
The algorithms handle event mixing, PMT response modeling, electronics response modeling, and trigger response modeling. Each algorithm is implemented using four tools dedicated to each sub-system: LPMT and SPMT systems in CD, the WP system, and the TT system.

The \texttt{EvtMixingTool} invoked by {\color{\mycolor}{\texttt{PMTSimAlg}}} combines different types of
physical events based on event rate inputs. Each physical event loaded
is assigned a timestamp which allows the detector simulation hit time information 
to be adjusted prior to mixing. Subsequently, the hits are unpacked from the 
loaded events and converted into pulses by the appropriate PMT tools, taking 
into account PMT effects such as gain resolution and variation, dark noise, 
pre-pulse and after-pulse, transit time, and transit time spread. 
Finally, all generated pulses are pushed into the pulse buffer in time order. 
This hit-level event mixing approach corresponds closely to the real-world situation, 
allowing \textcolor{\mycolor}{the} pileup of multiple event types to be handled naturally and correctly. 

The \texttt{TrigSimAlg} sub-system tools simulate the actions of the corresponding 
hardware trigger cards. Simulated triggers are modelled by generating a trigger signal 
that is pushed into the trigger buffer together with trigger information.  
The tools used in \texttt{ElecSimAlg} sub-system process the pulses within the readout time window,
applying various electronics effects such as waveform modelling and digitization 
and generating the outputs. Subsequently, EDM objects are created and filled by 
corresponding handler tools. 

\subsection{``Pull'' mode workflow}
\label{sec:pullmode}

As physical events may produce multiple readout events spanning hundreds of microseconds,  
or even longer, event mixing implementations must carefully manage memory resources. 
The ``pull'' mode avoids excessive memory usage with user configurable time ranges 
and a processing time window (PTW) which is longer than the readout time window.  
Only events within the PTW are loaded, unpacked and mixed.

\begin{figure}[htb]\centering
        \includegraphics[width=85mm]{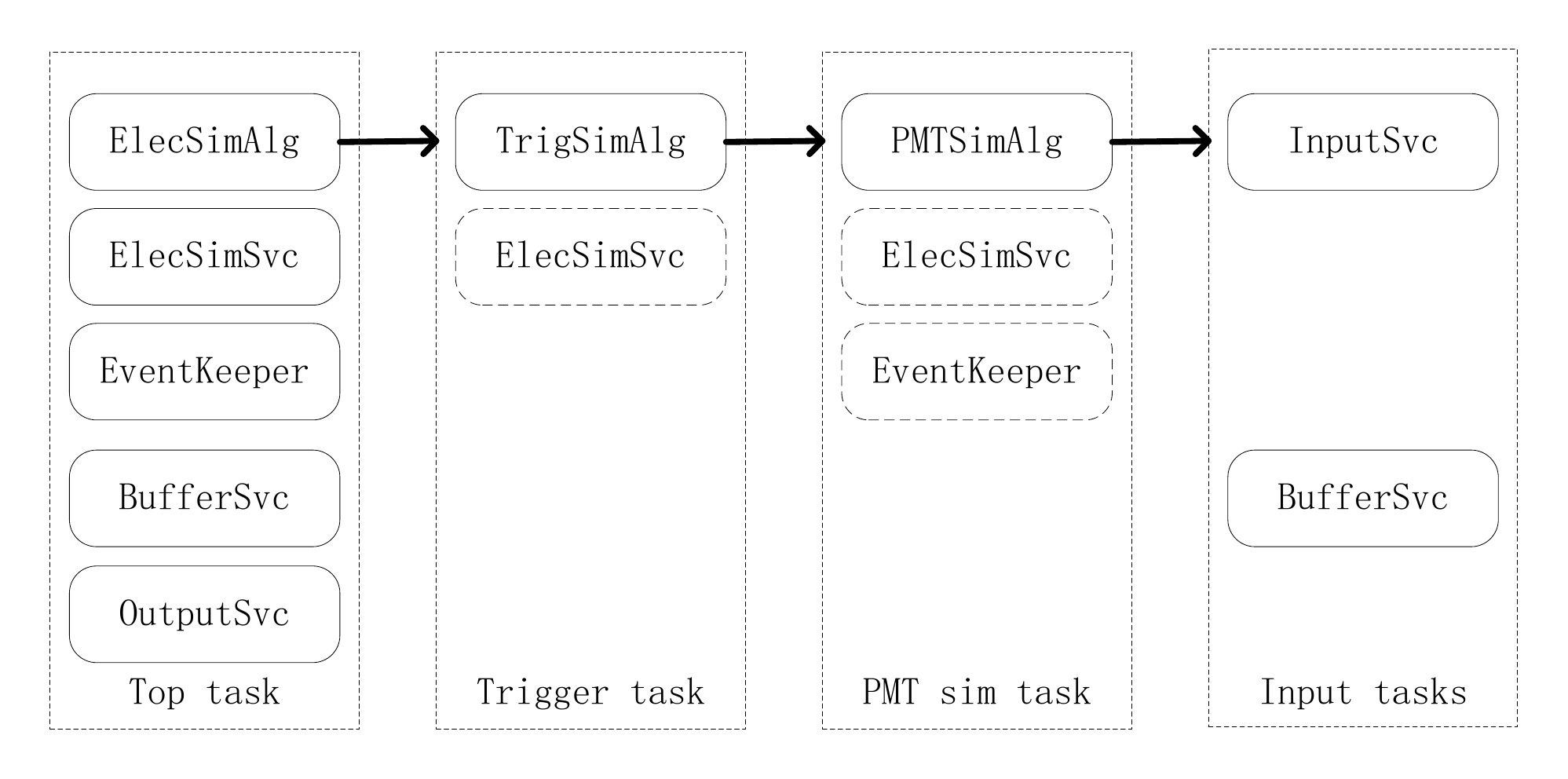}
        \caption{%
Pull mode electronics simulation workflow showing objects relevant to each task
with dashed boxes indicating shared objects. The arrows illustrate the control hierarchy
between the tasks with \texttt{ElecSimAlg} from the top task invoking \texttt{TrigSimAlg}
when the trigger buffer is empty. \texttt{TrigSimAlg} in turn invokes \texttt{PMTSimAlg} 
when the pulse buffer has insufficient pulses, which in turn invokes \texttt{InputSvc} 
to load the various event types.%
}
        \label{fig:ElecSimWorkFlow}
\end{figure}

In reality, PMT hits are triggered leading to pulses that are digitized into readout events. 
Pull mode mixing, as illustrated in Figure~\ref{fig:ElecSimWorkFlow}, operates in the reverse manner 
starting from \texttt{ElecSimAlg}, which creates readout events only when a trigger can be 
created from the simulated pulses. 
When the trigger buffer is empty \texttt{TrigSimAlg} looks for pulses to create a trigger. 
If there are insufficient pulses within the PTW to produce a trigger, the pulses within and 
before the PTW are deleted and a new PTW is opened and populated with pulses by 
the \texttt{PMTSimAlg} loading more events with timestamps inside the PTW.  
The \texttt{TrigSimAlg} continues to try to create a trigger from the updating 
pulse buffer until it succeeds allowing \texttt{ElecSimAlg} to create a readout event. 
Both pulse and trigger buffers operate in pipeline style with processed data removed 
and new data added. 

The implementation is organized using SNiPER roles, each algorithm is associated 
with a task that is only invoked when there is insufficient data in the corresponding buffer. 
Communication between \textcolor{\mycolor}{the} algorithm and task uses the SNiPER incident mechanism. 
When a task is invoked, registered algorithms and tools are executed once and 
tools registered in the algorithm are also invoked. This leads to 
all the relevant services of the tasks being updated. 

To mix events from different files, users need only assign labels and rates 
to the event types. The mixing implementation then proceeds as described above, 
writing events that \textcolor{\mycolor}{mimic} real data into the output buffer.   


\section{Data Model, MC Truth and Event Correlation}
\label{sec:evt_corr}

\subsection{Event data model and ROOT I/O}
\label{sec:EDM}

\begin{table*}[htb]
    \centering
    \caption{Event Data Models used by JUNO simulation}
    \label{table:edm}
    \begin{tabular}{|l|l|l|l|l|}
    \hline
    Stage      & Header object             & Event object           & Contained objects    & Default EDM path           \\
    \hline
    Generator  & \texttt{GenHeader}        & \texttt{GenEvt}        & tracks               & \texttt{/Event/Gen}        \\
    \hline
    Simulation & \texttt{SimHeader}        & \texttt{SimEvt}        & tracks and hits (CD/TT/WP)      & \texttt{/Event/Sim}        \\
    \hline
    Trigger    & \texttt{CdTriggerHeader}  & \texttt{CdTriggerEvt}  &                      & \texttt{/Event/CdTrigger}  \\
               & \texttt{WpTriggerHeader}  & \texttt{WpTriggerEvt}  &                      & \texttt{/Event/WpTrigger}  \\
               & \texttt{TtTriggerHeader}  & \texttt{TtTriggerEvt}  &                      & \texttt{/Event/TtTrigger}  \\
    \hline
    Readout    & \texttt{CdWaveformHeader} & \texttt{CdWaveformEvt} & waveforms (CD)       & \texttt{/Event/CdWaveform} \\
               & \texttt{CdLpmtElecHeader} & \texttt{CdLpmtElecEvt} & t/q pairs (CD LPMT)  & \texttt{/Event/CdLpmtElec} \\
               & \texttt{CdSpmtElecHeader} & \texttt{CdSpmtElecEvt} & t/q pairs (CD SPMT)  & \texttt{/Event/CdSpmtElec} \\
               & \texttt{WpElecHeader}     & \texttt{WpElecEvt}     & t/q pairs (WP)       & \texttt{/Event/WpElec}     \\
               & \texttt{WpWaveformHeader} & \texttt{WpWaveformEvt} & waveforms (WP)       & \texttt{/Event/WpWaveform} \\
               & \texttt{TtElecHeader}     & \texttt{TtElecEvt}     & t/q pairs (TT)       & \texttt{/Event/TtElec}     \\
    \hline
    \end{tabular}
\end{table*}

The JUNO event data model (EDM)~\cite{Li:2017zku} defines the content of the event
data objects consumed and produced by all the algorithms that constitute the 
simulation workflow, as shown in Table~\ref{table:edm}. The physics generator algorithm produces \texttt{GenEvt}
objects that are consumed by the detector simulation algorithm in order 
to produce \texttt{SimEvt} objects. Finally, the electronics simulation consumes 
\texttt{SimEvt} objects and produces \texttt{TriggerEvt} and \texttt{ElecEvt} objects.

All event data objects follow a two-level design with {\color{\mycolor}{lightweight}} header objects
that refer to event objects containing the bulk of the data. 
An event navigator consists of a list of header objects, which can be accessed by the 
corresponding EDM path. The header objects are loaded into memory first, subsequently, 
event objects are loaded only when needed reducing resource usage.   
The output of detector simulation stores both \texttt{GenEvt} and \texttt{SimEvt}. 
\texttt{GenEvt} objects refer to \texttt{HepMC::GenEvent}. 
ROOT dictionaries of HepMC objects are generated to facilitate access when persisted to ROOT files.
\texttt{SimEvt} objects hold collections of tracks, and hits in CD, WP and TT. 
The track class includes members such as particle type, positions, momenta 
and total deposited energy. Both primary and secondary tracks from Geant4 are stored.
Storage of secondaries is especially relevant with the neutron capture process and
radioactivity decay processes. Hits from the CD and WP share a common hit type, with 
a separate type used for TT hits. 

The outputs from the electronics simulation consist of triggers and readouts. 
The trigger types are divided into three classes: \texttt{CdTriggerEvt}, 
\texttt{WpTriggerEvt} and \texttt{TtTriggerEvt}. The readout types are divided
into six classes: \texttt{CdWaveformEvt} stores the waveforms of LPMTs in CD; 
\texttt{CdLpmtElecEvt} stores time and charge pairs of LPMTs in CD;
\texttt{CdSpmtElecEvt} stores time and charge pairs of SPMTs in CD;
\texttt{WpElecEvt} stores time and charge pairs in WP;
\texttt{WpWaveformEvt} stores the waveforms in WP;
\texttt{TtElecEvt} with time and charge pairs in TT. 
This design provides flexibility to match resource usage to requirements.

A ROOT-based I/O system is used to read and write event data model objects.
On reading an event from a file the input service first reads an event navigator object.
Loading of header and event objects is deferred until requested, avoiding unnecessary I/O.
When writing an event the output service writes the event navigator, header objects, and 
event objects. Only headers and event objects associated with the event navigator are written.

{\color{\mycolor}{In order to reduce memory usage in the detector simulation, a huge event is split into sub-events. All the information except the collection of PMT hits in CD are stored in the first sub-event, while the collection of PMT hits in \textcolor{\mycolor}{the} CD are split into the rest sub-events. The event navigators share the same information for these sub-events. Using a dedicated ROOT input service, these sub-events could be merged again according to the same information from event navigators.}}
Such features are configured with separate SNiPER sub-tasks that are invoked using an incident mechanism.

The loading of various event samples needed for event mixing is implemented 
using the SNiPER multiple sub-tasks feature. Each sub-task is configured with 
an input service and a buffer management service. When event mixing loads 
a sample the corresponding sub-task is invoked causing the input service 
to load the event into the data buffer providing access to the corresponding input task.  

\subsection{Monte Carlo truth}
\begin{figure}[htb]\centering
	\includegraphics[width=85mm]{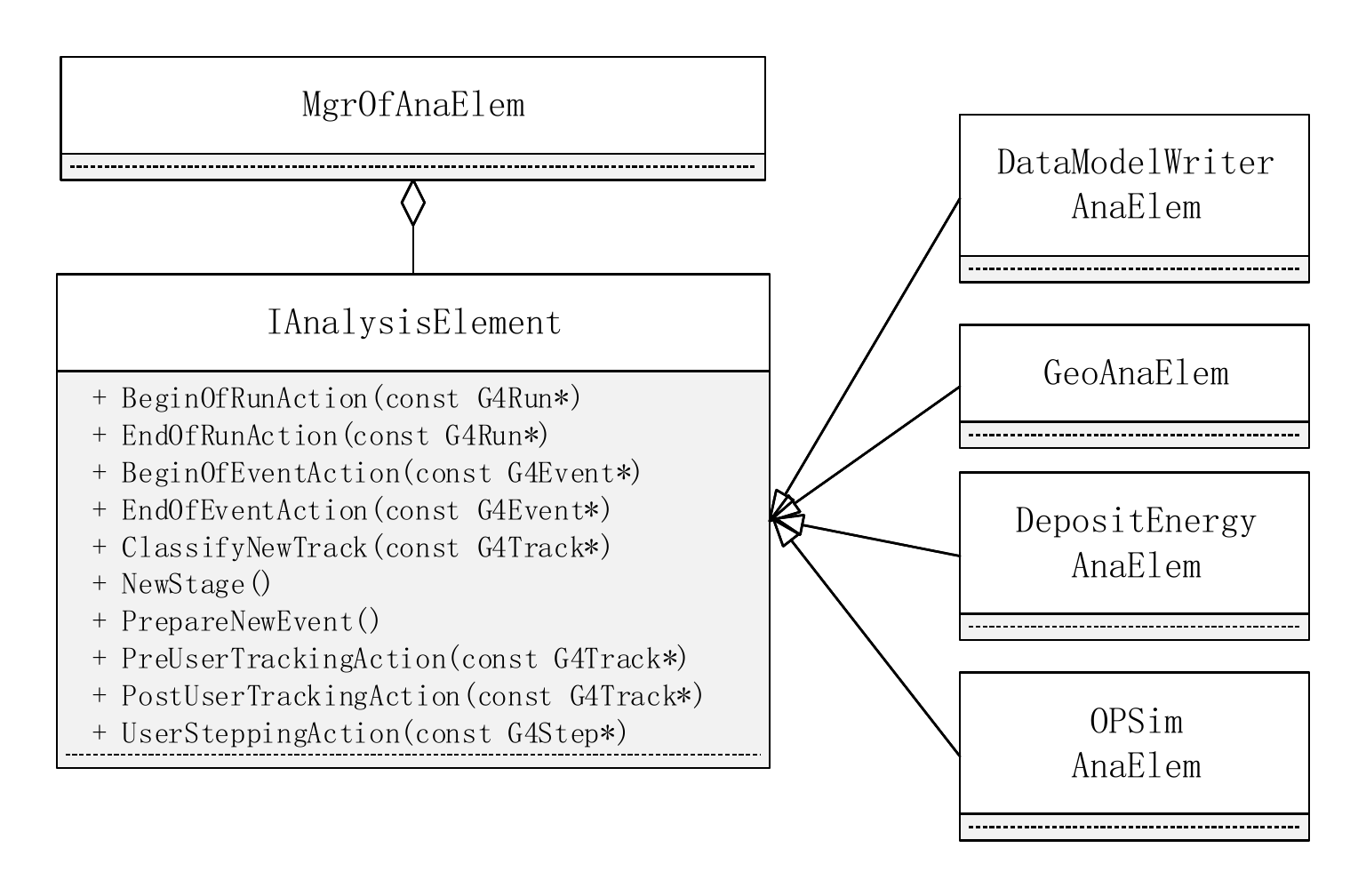}
	\caption{Geant4 user action access from multiple independent tools}
	\label{fig:UserAction}
\end{figure}

Geant4 user actions provide access to MC truth information via objects such as G4Event, G4Track, and G4Step. 
This access is shared with multiple independent tools, as illustrated in Figure~\ref{fig:UserAction}, 
using a tool manager that maintains a list of tools that implement the \texttt{IAnalysisElement} interface. 
The Geant4 user actions via the tool manager invoke the corresponding methods 
of all the tools according to their registration order.   

\subsection{Event correlation}

\begin{figure}[htb]\centering
	\includegraphics[width=85mm]{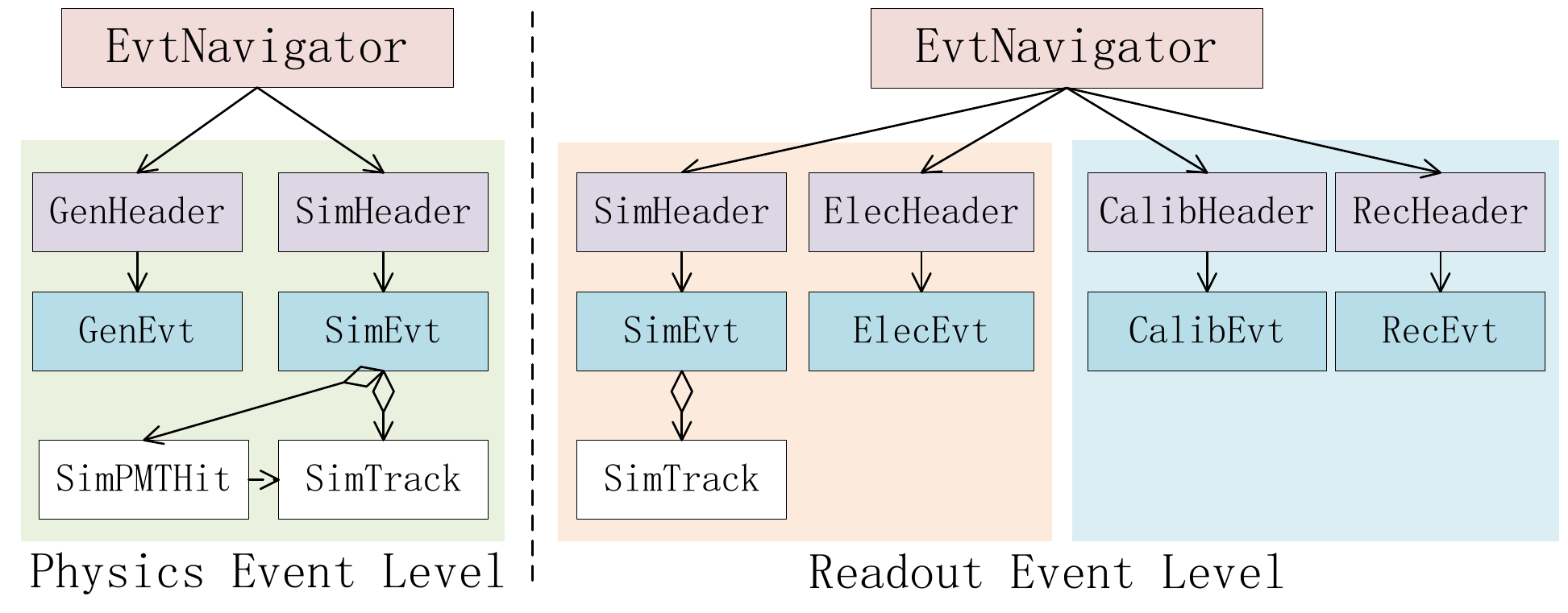}
	\caption{Event correlation handling}
	\label{fig:EventCorrelation}
\end{figure}

{\color{\mycolor}{Events at the physics event level, such as from the IBD process, 
may be split into two or more events at the readout event level
and also background events may be mixed in with signal events.
Figure~\ref{fig:EventCorrelation} shows how to handle such a case. }}
An {\tt EvtNavigator} object that maintains references to associated objects
allows the relationships between objects at different levels 
to be navigated, for example, relating readout objects at the electronics simulation level 
with primary particles at the detector simulation level. 
When a detector simulation event is split into two readout objects, 
{\color{\mycolor}{two {\tt EvtNavigator} objects at \textcolor{\mycolor}{readout} event level will be created
 and both maintain information concerning the primary stage,
 so the events after reconstruction could be correlated with the same \texttt{SimTrack} objects.}}
An {\tt EventKeeper} utility is provided that, using the relationship information recorded in the {\tt EvtNavigator} objects, 
is able to rebuild {\tt SimEvt} corresponding to readout events including all the track level information. 
The number of hits contributing to the readout is also provided, but hit objects are not duplicated. 

\section{Strategies to Improve Computing Performance}
\label{sec:fast_sim}
\subsection{Multi-threaded detector simulation}
JUNO simulation has been updated to enable multi-threaded running, 
profiting from Geant4 functionality that provides event-level parallelization.
Events are processed in different threads, while the detector geometry is shared.
For the JUNO geometry with many thousands of PMTs, this enables a significant reduction in memory resources.
The parallelized SNiPER framework is implemented using high-level task objects provided by  
the Intel Threading Building Blocks (TBB) library, avoiding low-level thread management. 
Each task object is configured with {\tt Algorithm} and {\tt Service}
components and then scheduled to run in a dedicated thread. 

Parallelized processing presents a challenge \textcolor{\mycolor}{in} handling events which must 
retain their time ordering, such as events from a supernova burst. 
A global buffer-based technique is developed~\cite{Zou:2019cyq} to maintain time ordering,
that orders generated objects within the global buffer prior to simulating each 
in dedicated threads. Finally completed events are saved by a ROOT I/O service 
from another dedicated thread~\cite{Lin:2019unc}.%


\subsection{Fast simulation}

The huge computational and memory demands of the full JUNO simulation of millions of optical photons 
has motivated \textcolor{\mycolor}{the} development of a voxel-based parameterization~\cite{Lin:2016vua}.  
Instead of propagating optical photons the parameterization samples
pre-generated distributions of the number of photoelectrons 
and hit times representing the PMT response.
At each step the appropriate distributions for the current position 
is accessed allowing random sampling to yield a number of photoelectrons 
and hit time for each PMT. 

\begin{figure}[htb]\centering
	\includegraphics[width=85mm]{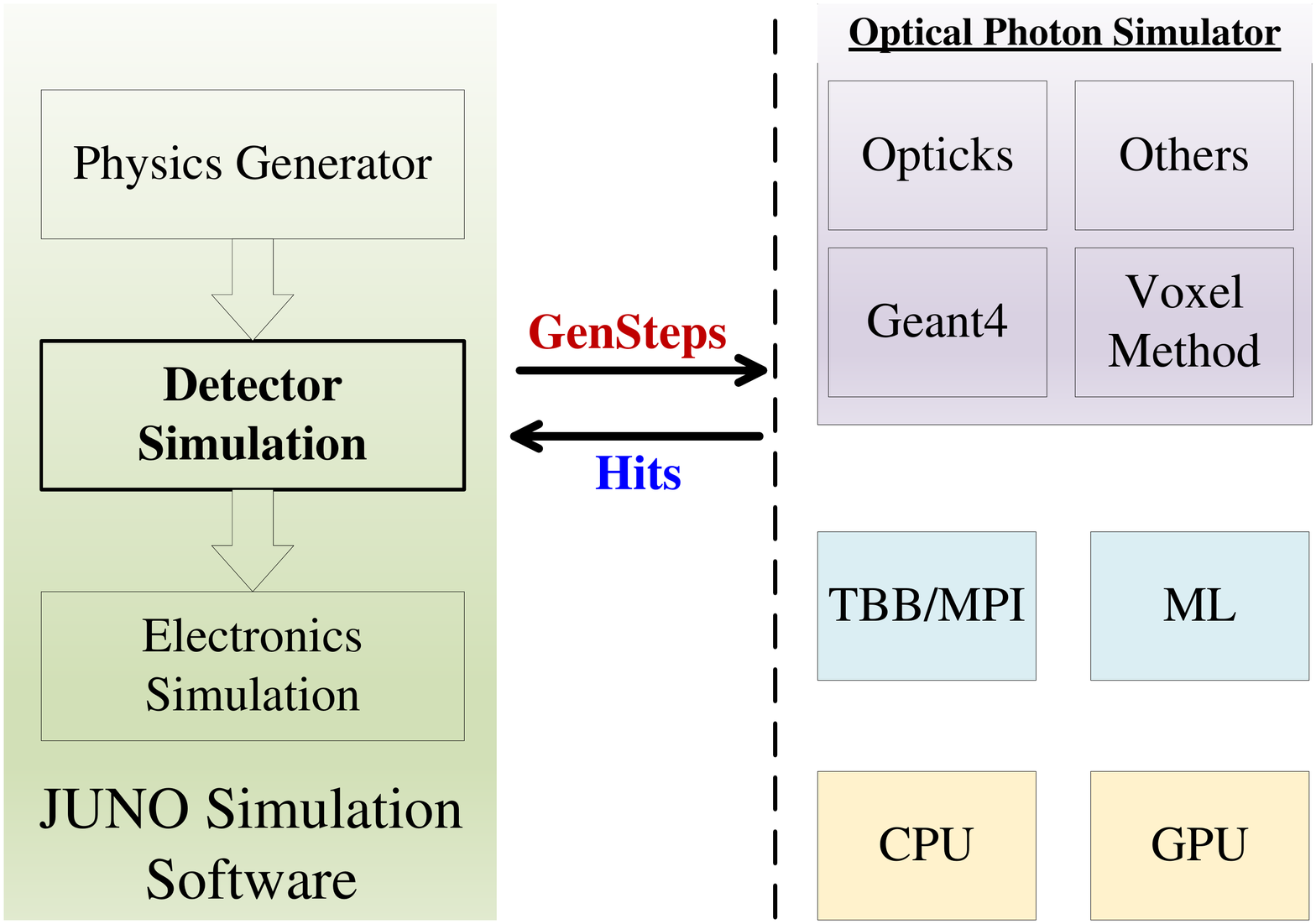}
	\caption{Deferred optical propagation method to integrate different optical photon simulators}
	\label{fig:deferred}
\end{figure}

Figure~\ref{fig:deferred} illustrates a unified approach to integrating 
the various optical photons simulators which allow the optical simulation 
to be deferred until after selection criteria have been satisfied~\cite{Lin:2022rxn}.
The simulators include the standard Geant4 optical photon simulator, the voxel method 
and the Opticks GPU-based simulator.
%
%


\subsection{Opticks GPU optical photon simulation}

Opticks{\color{\mycolor}{~\cite{Blyth:2021gam,opticksURL}}} performs a 
full optical photon simulation equivalent to the Geant4 optical photon simulation 
that benefits from the high performance ray tracing provided by NVIDIA GPUs. 
The Opticks implementation is based on the NVIDIA OptiX 7 API~\cite{10.1145/1778765.1778803}. 
All aspects of the Geant4 context relevant to optical photon generation and propagation 
such as the detector geometry, optical physics and the optical photons 
are translated into appropriate forms and are uploaded to the GPU. 
Detector geometry on the GPU is modelled with NVIDIA OptiX intersection, bounding box 
and closest hit programs and the buffers that these programs access. 
Opticks provides \textcolor{\mycolor}{an} automated translation of Geant4 geometries into these buffers, 
starting by traversing the Geant4 volume tree converting materials, surfaces, solids,
volumes and sensors into Opticks equivalents.

\begin{figure}[htb]\centering
	\includegraphics[width=70mm]{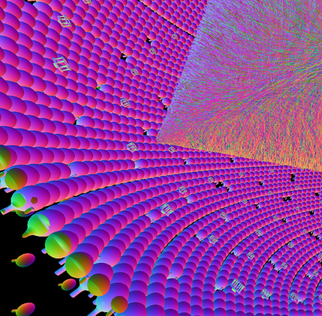}
	\caption{
   {\color{\mycolor}{This figure shows the Cerenkov and scintillation photons produced by a 200 GeV muon traveling across the JUNO detector. Each line represents a single optical photon and colors indicate the polarization direction. The gaps among PMTs are caused by the stainless steel supporting bars which are not shown in the picture. These bars are used to support the acrylic sphere.}}
}
	\label{fig:OpticksMuon}
\end{figure}

Photons are brought to the GPU via NVIDIA OptiX ray generation programs
which contain CUDA ports of the photon generation loops from Geant4 scintillation 
and Cerenkov processes. These programs together with {\color{\mycolor}{``Genstep''}} 
data structures which are collected from Geant4 on the CPU allow 
the photons to be generated on the GPU.
As Geant4 has no {\color{\mycolor}{``Genstep''}} interface it is necessary to modify the classes
representing scintillation and Cerenkov processes. Instead of generating
photon secondary tracks in a loop the {\color{\mycolor}{``Genstep''}} parameters such as the process
type code, the number of photons to generate and the line segment along which
to generate them and all other parameters used in the generation loop are
collected and uploaded to the GPU, typically at the end of the event.
This allows the photons to be allocated, generated and propagated entirely on the GPU, 
minimizing transfer overheads and allowing CPU memory usage to be restricted to optical photons 
that hit photomultiplier tubes, which are copied back to the CPU and added to Geant4 
hit collections, allowing the rest of the simulation to proceed unmodified. 
As illustrated in Figure~\ref{fig:OpticksMuon} Opticks also provides OpenGL GPU based 
visualizations of detector geometries and optical photon propagations. 

Offloading the computational and memory burdens of simulating millions
of optical photons to the GPU eliminates processing bottlenecks.
As the optical photons in JUNO can be considered to be produced only by the 
scintillation and Cerenkov processes and yield only hits on photomultiplier tubes 
it is straightforward to integrate the Opticks external optical photon 
simulation together with the Geant4 simulation of all other particles.
{\color{\mycolor}{Reference~\cite{Blyth:2021gam} provides further details on Opticks and its integration 
with the JUNO simulation framework and potential performance benefits. }}


\section{Visualization}
Visualizations of detector geometries and  
of actual and simulated events provide the most effective way to 
communicate the principals of detector operation to students and the general public.
Interactive display interfaces that present particle 
trajectories, physics processes and hits on detector elements  
can greatly assist understanding and prove insightful 
for the development of reconstruction and calibration approaches.
In addition, visualization is vital for the development of the simulation
allowing geometry problems such as overlaps to be identified.

\begin{figure}[htb]\centering
	\includegraphics[width=85mm]{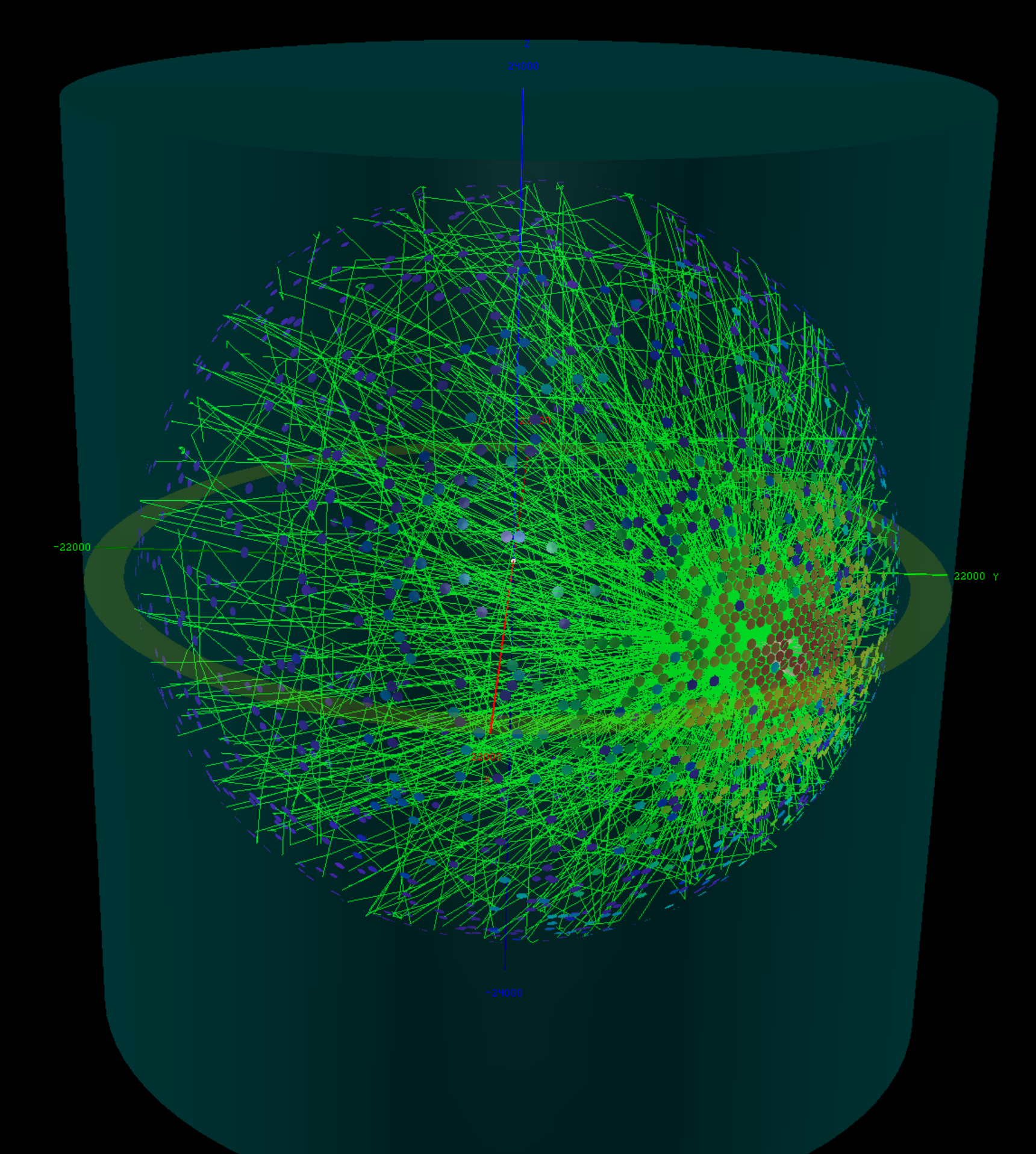}
	\caption{SERENA display of particle trajectories in a simulated event.}
	\label{fig:VisTruth}
\end{figure}
%

\begin{figure}[htb]\centering
	\includegraphics[width=85mm]{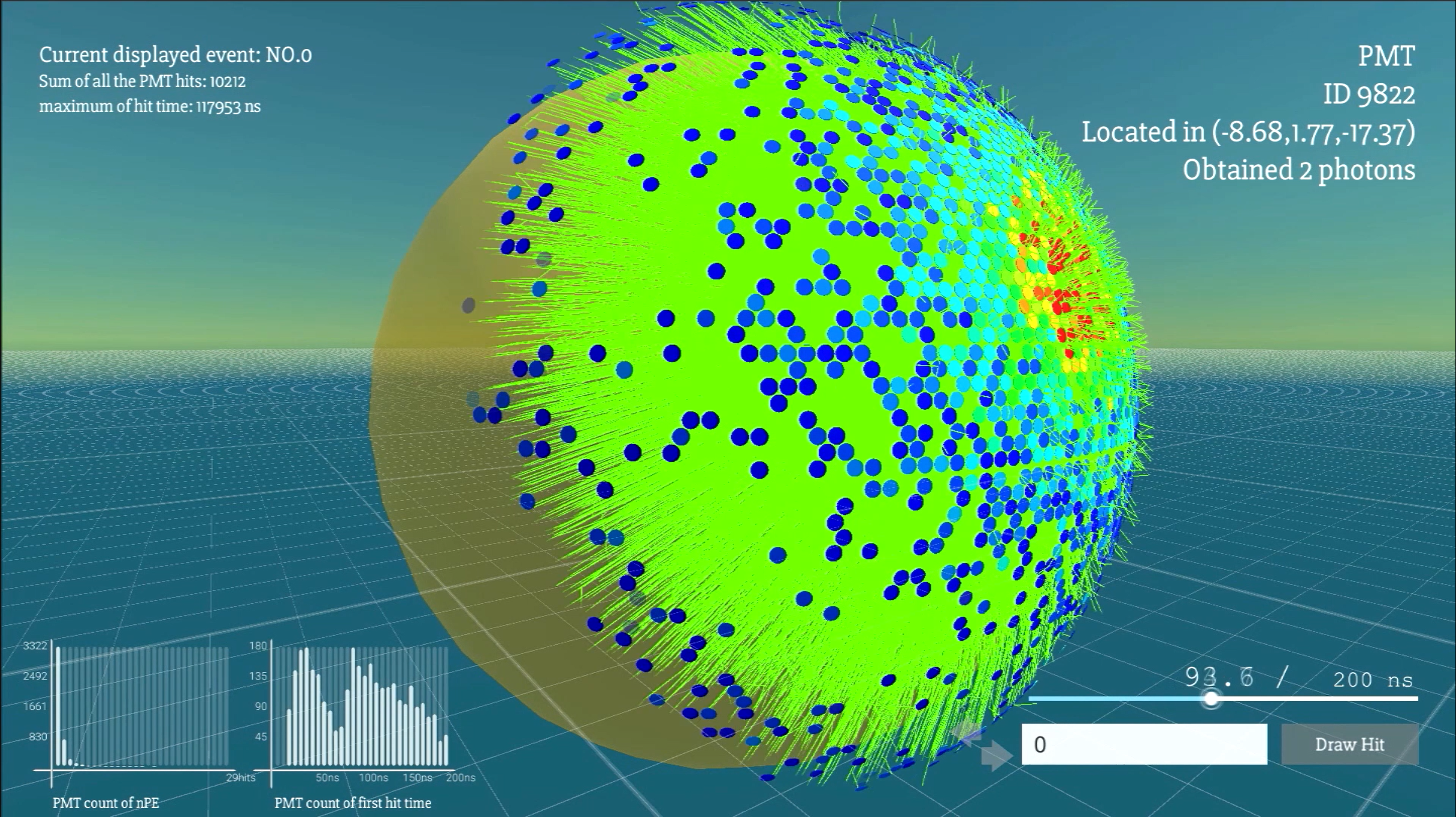}
	\caption{ELAINA display of particle trajectories in a simulated event.}
	\label{fig:ELAINA}
\end{figure}
%


SERENA~\cite{You:2017zfr} and ELAINA~\cite{Zhu:2018mzu} are two independent event display applications developed for JUNO. 
SERENA is a ROOT based event display package that is integrated with the offline software
geometry service~\cite{Li:2018fny,Huang:2022} and event data model. 
By reading the output files from the simulation program, the trajectories
and hits in \textcolor{\mycolor}{an} MC event can be visualized in the event display package, as shown
in Figure~\ref{fig:VisTruth} and \ref{fig:ELAINA}.
Users also have the flexibility to filter or highlight different physical
processes or optical processes, such as Fresnel refraction, total internal reflection,
or Rayleigh scattering. The interface allows \textcolor{\mycolor}{the} selection of specific trajectories or hits
for which to display further information.  
The other event display application ELAINA is an Unity~\cite{unity} based program 
which is independent of the offline software, facilitating installation on
a wide range of operating systems and providing eye-catching visual effects. 
%
%
%


\section{Summary}
Simulation software is an essential component of the JUNO experiment, 
playing an important role across detector design and commissioning,
offline data processing and physics analysis. The simulation software 
must be well designed and implemented and operate with reasonable
computing resources in order to meet the needs of the experiment 
throughout its life time of more than 30 years. 
The broad energy range of JUNO's physics program from keV to GeV,
and the scale of the JUNO detector make it challenging to achieve this goal. 
This paper has presented the design and implementation of the JUNO simulation software, 
comprising the four building blocks of generator, detector simulation, event
mixing and electronics simulation. The software is developed in C++ and Python 
and is based on the SNiPER framework and depends on several external
libraries, including Geant4, CLHEP, ROOT and BOOST. 
A {\color{\mycolor}{modular}} design style has been used, based on SNiPER dynamically loadable elements DLEs, 
enhancing maintainability, extensibility and reliability. 

Key challenges of the simulation have been highlighted together with 
strategies to address them. 
A unified and flexible generator interface has been developed that 
facilitates \textcolor{\mycolor}{the} integration of diverse physics generators and dynamic deployment of 
calibration sources. 
Flexible parameter access from a variety of data sources, such as files and databases, 
is provided via a single consistent interface used from all processing stages.

Several approaches to improve the computing performance of high energy event simulation 
are implemented, for example reducing memory consumption with multi-threading, merging hits 
and event splitting or using a parameterized fast simulation. 
Integration with Opticks speeds up the simulation and reduces CPU memory consumption 
by offloading the optical photon simulation to the GPU. 

Time correlated experimental data is mimicked using hit-level event mixing 
implemented with a ``pull'' mode that limits memory consumption.
Machinery that maintains object relationships despite splits and merges arising from 
the simulation and utilities giving complete Monte Carlo truth information for 
readout events are provided.  

The JUNO simulation software has been successfully used to perform large scale Monte Carlo
productions for the JUNO collaboration. Attaining the goal of providing a precise reproduction 
of experimental data across a broad range of physics analyses in JUNO 
requires a long-term commitment to ongoing development that strives to continually 
improve our understanding of the JUNO detector and the performance of the simulation. 


\begin{acknowledgements}
We gratefully acknowledge support from the Strategic Priority Research Program of the Chinese Academy of Sciences (CAS), Grant No. XDA10010900. This work is also supported in part by National Natural Science Foundation of China (NSFC) under grant No. 11875279, No. 12275293, and No. 12025502, supported in part by Youth Innovation Promotion Association, CAS. This work has been partially supported by CNPq, grant No.407149/2021-0.
\end{acknowledgements}

\bibliographystyle{spphys}  
\bibliography{mybibfile}

\end{document}